\documentclass[%
 aip,
 amsmath,amssymb,
 reprint,%
]{revtex4-1}

\usepackage{graphicx}% Include figure files
\usepackage{dcolumn}% Align table columns on decimal point
\usepackage{bm}% bold math
\usepackage[utf8]{inputenc}
\usepackage[T1]{fontenc}
\usepackage{mathptmx}
\usepackage{etoolbox}

\makeatletter
\def\@email#1#2{%
 \endgroup
 \patchcmd{\titleblock@produce}
  {\frontmatter@RRAPformat}
  {\frontmatter@RRAPformat{\produce@RRAP{*#1\href{mailto:#2}{#2}}}\frontmatter@RRAPformat}
  {}{}
}%
\makeatother
\begin{document}

\preprint{AIP/123-QED}

\title{Critical Phenomena and Strategy Ordering with Hub Centrality Approach in the Aspiration-based Coordination Game}
% Force line breaks with \\
\author{Wonhee Jeong}
\author{Unjong Yu}%
 \email{uyu@gist.ac.kr}
\affiliation{%
   Department of Physics and Photon Science, Gwangju Institute of Science and Technology, Gwangju 61005, South Korea
}%

\date{\today}% It is always \today, today,

\begin{abstract}
We study the coordination game with an aspiration-driven update rule in regular graphs and scale-free networks. We prove that the model coincides exactly with the Ising model and shows a phase transition at the critical selection noise when the aspiration level is zero. It is found that the critical selection noise decreases with clustering in random regular graphs. With a non-zero aspiration level, the model also exhibits a phase transition as long as the aspiration level is smaller than the degree of graphs. We also show that the critical exponents are independent of clustering and aspiration level to confirm that the coordination game belongs to the Ising universality class.
As for scale-free networks, the effect of aspiration level on the order parameter at a low selection noise is examined. In model networks (Barab\'{a}si-Albert network and Holme-Kim network), the order parameter abruptly decreases when the aspiration level is the same as the average degree of the network. In real-world networks, in contrast, the order parameter decreases gradually. We explain this difference by proposing the concepts of hub centrality and local hub. The histogram of hub centrality of real-world networks separates into two parts unlike model networks, and local hubs exist only in real-world networks. We conclude that the difference of network structures in model and real-world networks induces qualitatively different behavior in the coordination game.
\end{abstract}

\maketitle

\begin{quotation}
The coordination game describes the emergence of standards, opinion formation, and the diffusion of innovations, which are important problems in sociology, business administration, and statistical mechanics. In this paper, we study the coordination game with an aspiration-based update rule, where each agent prefers to change its strategy if the payoff is small relative to the aspiration. For regular graphs, we demonstrate that the game shows phase transition into a frozen state dominated by one strategy as the selection noise decreases. Interestingly, the critical exponents, which determine the critical behavior at the transition, are the same as the Ising model in spite of the non-equilibrium nature of the coordination game. As for scale-free networks, we discovered a striking difference between model networks and real-world networks in the dynamics of the coordination game. We show the difference is from local hubs, which exist only in real-world networks. Local hubs play dominant roles within their neighborhoods, though their degrees are not so large. The local hubs are identified by the hub centrality, which we propose in this paper. We further discuss the dynamics of local hubs and relation with their neighbors.
\end{quotation}

\section{\label{sec:level1}Introduction\protect}
Since the proposal of Smith and Price,\cite{smith1973} the evolutionary game theory has received great attention and made significant progress in various disciplines.\cite{nowak2006five,RevModPhys.81.591,RevModPhys.91.045004} The evolutionary game theory is composed of three main parts: the payoff matrix of the game, the strategy update rule, and the network structure,\cite{PhysRevE.58.69,nowak1992evolutionary} which are controlled to study various phenomena. The payoff matrix determines the payoff of each agent depending on the strategies of the agent and its neighbors; a higher payoff helps the survival and diffusion of the strategy. The prisoner's dilemma game describes situations with a social dilemma, which hinders cooperative society. The mechanism that promotes the evolution of cooperation even with the social dilemma is one of the key questions in the evolutionary game theory.\cite{Pennisi05,nowak2006five,tanimoto2015fundamentals,PhysRevLett.97.258103,Fu2021evolution} On the other hand, the coordination game is to study the emergence of standards, the dynamics of opinion formation, or the diffusion of innovations.\cite{XU2021110380,weibull1995evolutionary,PhysRevE.93.052108,PhysRevE.95.012303,PhysRevE.96.042101,Jin21} Interestingly, the coordination game has some points common to spin models such as the Ising model and similar kinds of phase transitions are observed in coordination games.\cite{PhysRevE.93.052108,PhysRevE.95.012303,PhysRevE.96.042101}

The strategy update rule describes how agents keep or change their strategies in accordance with the payoff of themselves and their neighbors.\cite{nowak2006evolutionary,PhysRevE.80.026108,PhysRevE.94.032317,PhysRevE.77.017103,PhysRevE.101.062309} There are many types of strategy update rules\cite{SZABO200797}: imitation (adaption of a neighbor's strategy), reproduction (enforcement of the strategy to a neighbor), aspiration-based update, etc. The aspiration-based update rule~\cite{PhysRevE.94.032317,PhysRevE.77.017103,PhysRevE.101.062309} is the win-stay-lose-shift rule~\cite{Nowak93} in structured networks: when the payoff of an agent satisfies the aspiration level, the agent tends to keep its strategy; otherwise, it prefers to change its strategy.\cite{PhysRevE.94.032317,PhysRevE.77.017103,PhysRevE.101.062309} This update rule promotes cooperation relative to the other strategy update rules in the prisoner's dilemma game when the temptation of the defective strategy is high.\cite{PhysRevE.94.032317,PhysRevE.77.017103,PhysRevE.101.062309,Liu11}

Recently, the importance of the network structure is emphasized in various aspects.\cite{PhysRevLett.95.098104,PhysRevE.68.030901,mcavoy2020social,alvarez2021evolutionary,Santos3490,RevModPhys.81.591} The network describes the social relation between agents. Many kinds of social networks in the real world have been uncovered; but for a systematic study with controlled parameters, many model networks have been proposed.\cite{watts1998collective,Barabasi509,PhysRevLett.89.208701,PhysRevE.65.026107,Bollobas01,Jeong_2019} A model network can be generated for a given degree distribution, clustering coefficient, average shortest path length, and assortativity in most cases. However, it is not clear yet whether the model network behaves in the same way as the real-world network in each of the specific problems.\cite{PhysRevLett.89.208701,broido2019scale,holme2019rare} In the evolutionary game theory, the network structure is also important.\cite{mcavoy2020social,alvarez2021evolutionary,Santos3490,RevModPhys.81.591} The number of neighbors of an agent is critical to the payoff of the agent. Usually, an agent with more neighbors is assumed to have a higher probability to gain a large payoff. In the prisoner's dilemma game in a scale-free network, for instance, some agents have very large numbers of neighbors~\cite{Barabasi509} and they would obtain more payoff compared to other agents with fewer neighbors. Agents with many neighbors tend to be cooperative because cooperation outperforms defection on average for random neighbors, and they are advantageous in strategy diffusion; therefore, highly inhomogeneous networks are favorable to the evolution of cooperation.\cite{PhysRevLett.95.098104} For this reason, although the level of cooperation is influenced by many factors (noise level, dynamic rules, payoff parameters, etc), the detailed network structure is also crucial.\cite{Jeong_2019,CHEN2007379,PhysRevE.72.047107,PhysRevE.82.047101}

In this paper, we study the aspiration-based coordination game in various networks including real-world networks. We show that the game is mapped to the Ising model in a specific condition. We investigate the phase transition in random regular networks with varying clustering. We also examine the effect of the aspiration level on critical selection noise and critical exponents. Finally, we compare the results of the aspiration-based coordination game in model networks and real-world networks with scale-free degree distribution. We discovered qualitatively very different behaviors and propose a new parameter, hub centrality to explain these. Based on the hub centrality, we define the local hub and discuss its role in the dynamics of the coordination game.

In the following section, we introduce the model and method of our work. In section~\ref{sec:level3}, we present results. We show the coordination game with the zero aspiration level is the same as the Ising model, which is also verified numerically. And we show the results of the non-zero aspiration level cases. In subsection~\ref{sec:level3-4}, we compare the results of model networks and real-world networks. The difference between them is explained using the concept of the local hub. Finally, section~\ref{sec:level4} concludes.

\section{\label{sec:level2}Model and Method}
We consider various graphs and networks: square lattice, random regular graph (RRG),\cite{Bollobas01,Jeong_2019} Barab\'{a}si-Albert (BA) scale-free network,\cite{Barabasi509} Holme-Kim (HK) scale-free network,\cite{PhysRevE.65.026107} and real-world networks.\cite{nr} These graphs and networks consist of nodes and edges. Nodes represent agents and the interactions between agents are through edges. When two agents are connected by an edge, they are neighbors of each other. The number of neighbors of an agent $i$ is called the degree ($k_i$) of the agent.

In this paper, we study the coordination game with the aspiration-based update rule. There are two strategies in the game: strategy 1 (S1) and strategy 2 (S2). If the two agents who play the game have the same strategy, each of them gets a payoff of $2$. Otherwise, each of them gets a payoff of $-2$. Every agent plays the game with all of its neighbors and the payoff is accumulated. In the next section, we show that this payoff matrix makes the game equivalent to the Ising model in a specific condition.\cite{weibull1995evolutionary}

The aspiration-based strategy update rule is represented by the probability of strategy switch of
\begin{eqnarray}
 P_{i} = \frac{1}{1+\exp\left[(\pi_i - A)/\kappa \right]} , \label{eqn1}
\end{eqnarray}
where $\pi_i$ is the payoff of agent $i$. If the payoff of agent $i$ is less than the aspiration level $A$, the agent has more probability to change its strategy ($P_i>0.5$); otherwise, with higher probability the agent keeps its strategy. The selection noise (inverse selection strength) $\kappa$ characterizes the irrationality level in the decision process~\cite{PhysRevE.77.017103} and plays the same role as the temperature of the Ising model.\cite{plischke2006equilibrium,PhysRevE.58.69} For $\kappa \rightarrow \infty$, the agent changes its strategy with the probability of $1/2$. In the other limit of $\kappa \ll 1$, the agent changes its strategy for $\pi_i < A$ and keeps it for $\pi_i > A$; exactly at the point of $\pi_i = A$, the probability of strategy switch is $1/2$.

The strategy update is carried out asynchronously. Among $N$ agents, one agent is selected uniformly at random. The agent plays the games with all of its neighbors and accumulates payoff. Based on the accumulated payoff, the strategy switch probability is calculated. According to the probability, the agent changes or keeps its strategy. One generation is defined as $N$ such trials so that each agent has one chance of strategy update on average. We performed the simulation of at least $10\,000$ generations for the system to reach an equilibrium state. After an equilibrium state is achieved, we measured physical quantities for the system.

The order parameter in the coordination game is given by
\begin{eqnarray}
\langle m \rangle =  \left\langle \frac{\left|N_{\mathrm{S1}} - N_{\mathrm{S2}}\right|}{N} \right\rangle, \label{eqn3}
\end{eqnarray}
where $N$ is the number of all agents of the given network, $N_{S1}$ ({\it resp.} $N_{S2}$) is the number of agents with strategy S1 ({\it resp.} S2), and $\langle \cdots \rangle$ means the ensemble average.
It shows the degree of consensus between the two strategies; if all agents have the same strategy, $\langle m\rangle=1$ and it vanishes in a random strategy case. Based on Eq.~(\ref{eqn3}), susceptibility $\chi$ and Binder cumulant $U$ are defined as
\begin{eqnarray}
\chi &=& N \left(\langle m^2 \rangle - \langle m \rangle ^2 \right) \label{eqn4}\\
U &=& 1 - \frac{\langle m^4 \rangle}{3 \langle m^2 \rangle ^2} \label{eqn5} .
\end{eqnarray}
It is worth mentioning that energy and specific heat are not well-defined in this system.

\section{\label{sec:level3}Results and discussions}

\subsection{\label{sec:level3-1}Correspondence between the Ising model and the coordination game}

The Ising model~\cite{Ising} is represented as the Hamiltonian $H=-J\sum_{\langle i,j\rangle} S_i S_j$, where the spin $S_i$ at node $i$ may take on the values $\pm 1$. The exchange interaction parameter $J$ is set to $1$ and used as the energy unit. It has positive critical temperature $T_c$, at which long-range order emerges, in two and higher dimensions.\cite{Onsager,newman1999monte,plischke2006equilibrium} The critical behavior close to $T_c$ is described by the critical exponents, which are exactly known only in two dimensions.\cite{Onsager,newman1999monte,plischke2006equilibrium} The Ising model can be studied using the Markov-chain Monte Carlo method, where a spin is chosen uniformly at random and the spin is flipped according to a spin-flip probability $P(\mu \rightarrow \nu)$. Any function can be used as the spin-flip probability if it satisfies the detailed balance.\cite{newman1999monte,plischke2006equilibrium} Among many such functions, we may adopt the heat-bath algorithm:
\begin{eqnarray}
P(\mu \rightarrow \nu) = \frac{1}{1+\exp\left[(E_\nu - E_\mu)/{k_{B}T} \right]}, \label{eqn2}
\end{eqnarray}
where $E_\nu$ is the energy of state $\nu$ and $T$ is temperature. The Boltzmann constant $k_B$ is set to $1$ without loss of generality. The energy change ($E_\nu - E_\mu$) by the spin-flip is determined by the spin configuration of the neighbors of the chosen spin: $E_\nu - E_\mu = 2 D_s$, where $D_s$ is the number of neighbors with the same spin direction subtracted by the number of neighbors with the opposite spin direction before the spin-flip. Note that $2D_s$ is the same as the payoff in the coordination game studied in this paper, and so the spin-flip probability of Eq.~(\ref{eqn2}) is the same as the strategy switch probability of Eq.~(\ref{eqn1}) for $A=0$ if $k_BT$ is replaced by $\kappa$. The probability corresponds to Glauber dynamics of kinetic Ising model and potential game with Boltzmann-Gibbs distribution.~\cite{Glauber294,BLUME1993387} The logic rule for two strategies in fictitious game is also conceptually equivalent.~\cite{FUDENBERG1998631,MONDERER1996258,koopmans1951activity} Therefore, for the specific payoff matrix and aspiration level, the aspiration-based coordination model shows the same results as the Ising model in the critical selection noise (critical temperature), critical exponents, etc.

\subsection{\label{sec:level3-2}Critical behaviors of the coordination game with the zero aspiration level}
\begin{figure}[t]
\centering
\includegraphics[angle=270,width=1\columnwidth]{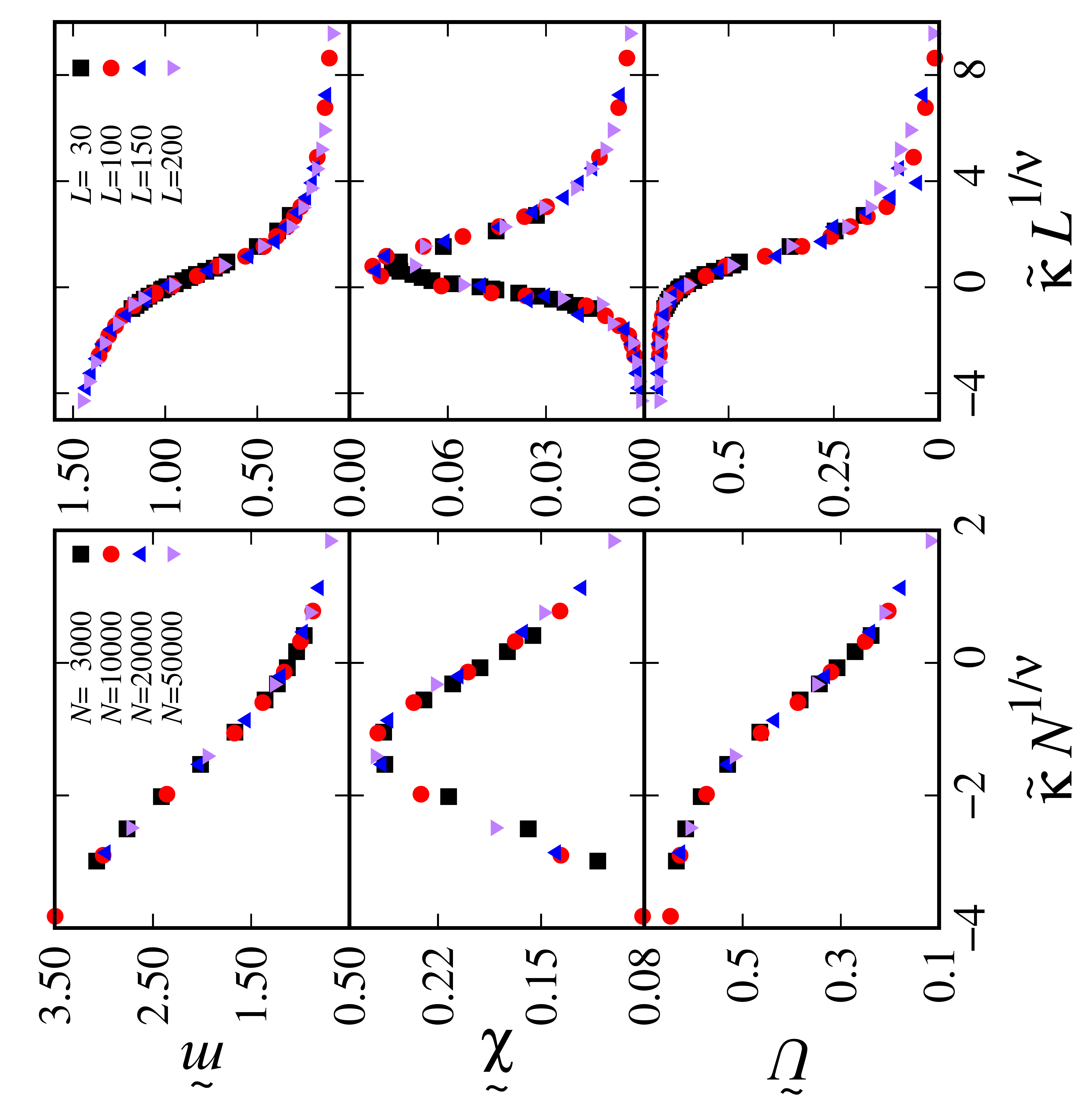}
\caption{Scaling functions ($\tilde{m}$, $\tilde{\chi}$, and $\tilde{U}$) as a function of the rescaled selection noise $\tilde{\kappa}$ for the aspiration-based coordination game in the RRG (the left panel) and in the square lattice (the right panel). The aspiration level is zero. The degree of each node of the RRG is four. Values of the critical selection noise and critical exponents are in Table~\protect\ref{tab1}.
Error bars are smaller than the symbol size.}
\label{fig1}
\end{figure}

In the previous subsection, we showed that the aspiration-based coordination game is equal to the Ising model for $A=0$, and it should have the phase transition at finite selection noise. We study the phase transition in the square lattice and RRG in this subsection. The RRG was generated by the algorithm introduced in Ref.~\onlinecite{Jeong_2019}. The results are averaged over $100\,000$ generations after the system reaches equilibrium and the results of $30$ independent simulations are averaged for the final results.

For each graph, we calculated the critical selection noise ($\kappa_c$) and three exponents ($\nu$, $\beta$, and $\gamma$) using finite-size scaling. Physical quantities near the critical selection noise are described by critical exponents:~\cite{PhysRevLett.47.693,PhysRevB.44.5081,PhysRevE.95.012101,newman1999monte,plischke2006equilibrium}
\begin{eqnarray}
m_L (\tilde{\kappa}) &=& L^{-\beta / \nu} \tilde{m} ( \tilde{\kappa} L^{1/\nu} ) \label{eqn6}\\
\chi_L (\tilde{\kappa}) &=& L^{\gamma/\nu} \tilde{\chi} ( \tilde{\kappa} L^{1/\nu} ) \label{eqn7}\\
U_L (\tilde{\kappa}) &=& \tilde{U} ( \tilde{\kappa} L^{1/\nu} ) \label{eqn8}\\
\tilde{\kappa} &=& \frac{\kappa - \kappa_{c}}{\kappa_{c}}, \label{eqn9}
\end{eqnarray}
where $L$ is the linear size of the lattice ($L=N^{1/2}$ for the square lattice). In the case of complex networks such as RRG, $L$ should be replaced by the number of agents $N$ in Eqs.~(\ref{eqn6})-(\ref{eqn8}). Since scaling functions have the same value regardless of the graph size at $\kappa = \kappa_{c}$, critical selection noise $\kappa_{c}$ and critical exponents can be calculated from the values of physical quantities at $\kappa = \kappa_{c}$.\cite{PhysRevE.95.012101,newman1999monte,plischke2006equilibrium} According to Eq.~(\ref{eqn8}), $\kappa_{c}$ can be obtained from the intersection of Binder cumulants of different graph sizes. After $\kappa_{c}$ is determined, the critical exponents can be calculated from Eqs.~(\ref{eqn6}), (\ref{eqn7}), and differentiation of Eq.~(\ref{eqn8}).

\begin{table}[b]
\caption{Critical selection noise $\kappa_c$ and critical exponents of the aspiration-based coordination game in the two-dimensional square lattice (SL), SL with $A=-4$, random regular graph (RRG) with $A=0$, highly clustered RRG with $A=0$, and RRG with $A=-4$. 
The degree $k$ is $4$ for all the graphs in this table.
Exact values of critical temperature ($T_c$) and critical exponents of the Ising model in the SL~\protect\cite{Onsager} and in the RRG are also presented for comparison. The exact values of $T_c=2/\ln[k/(k-2)]$ and critical exponents of the Ising model in RRG can be obtained by the Bethe-Peierls approach~\cite{Bethe,Peierls} and the mean-field theory.\protect\cite{plischke2006equilibrium}
}
\label{tab1}
\begin{tabular}{|l|l|l|l|l|}
\hline
\multicolumn{1}{|c|}{Graphs} & \multicolumn{1}{c|}{$\kappa_{c}$ ($T_c$)} & \multicolumn{1}{c|}{$1/\nu$} & \multicolumn{1}{c|}{$\gamma/\nu$} & \multicolumn{1}{c|}{$\beta/\nu$}  \\ \hline
SL               & $2.2688(4)$ & $0.96(6)$ & $1.76(1)$  & $0.11(1)$ \\ \hline
SL ($A=-4$)      & $3.292(1)$  & $0.99(5)$ & $1.756(4)$ & $0.10(2)$ \\ \hline
RRG              & $2.883(2)$  & $0.53(3)$ & $0.50(1)$  & $0.23(3)$ \\ \hline
RRG ($c=0.2034$) & $2.666(2)$  & $0.50(1)$ & $0.50(1)$  & $0.23(4)$ \\ \hline
RRG ($A=-4$)     & $4.050(1)$  & $0.50(3)$ & $0.50(1)$  & $0.24(1)$ \\ \hline
Ising (SL)       & $2 / \ln(1+\sqrt{2})$ & $1$ & $7/4$  & $1/8$     \\ \hline
Ising (RRG)      & $2 / \ln(2)$ & $1/2$     & $1/2$     & $1/4$     \\ \hline
\end{tabular}
\end{table}

Values of the critical selection noise and critical exponents calculated in this way are presented in Table~\ref{tab1}. The results for the square lattice are equal to the exact solutions of the two-dimensional (2D) Ising model~\cite{Onsager} as expected within the margin of error. As for the RRG, the critical exponents are consistent with the mean-field theory (MFT) due to small average-path-length.\cite{Barrat,PhysRevE.66.018101} Furthermore, we show that the scaling functions ($\tilde{m}$, $\tilde{\chi}$, and $\tilde{U}$) collapse independently of graph size in Fig.~\ref{fig1}. Due to the critical slowing down,\cite{newman1999monte,plischke2006equilibrium} which appears close to $\kappa_c$ in a local update algorithm, a somewhat large statistical error is unavoidable.

\begin{figure}[t]
\centering
\includegraphics[angle=270,width=1\columnwidth]{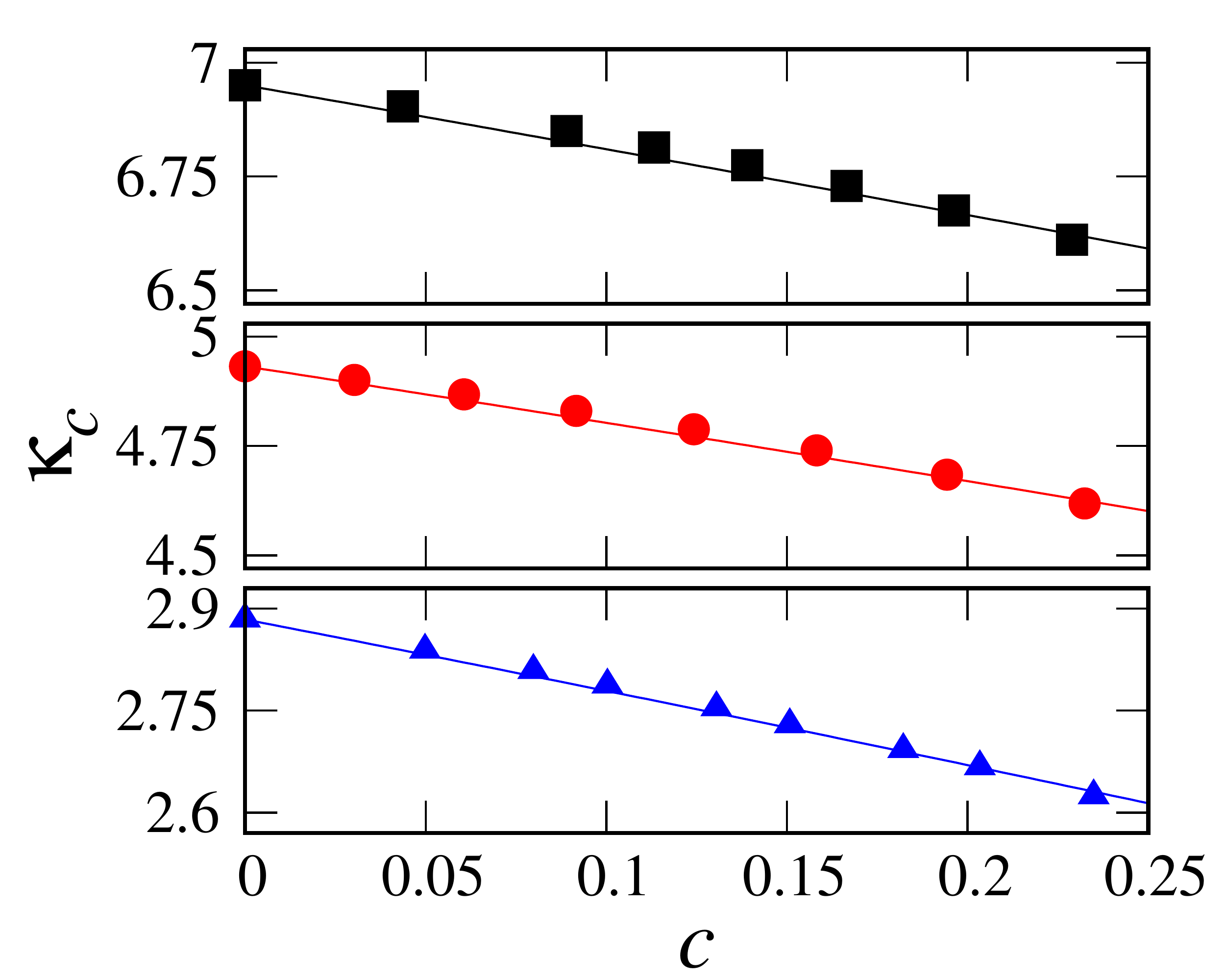}
\caption{The critical selection noise $\kappa_c$ of highly clustered RRGs as a function of clustering coefficient $c$. 
Symbols represent simulation results and solid lines are from the conjecture of Eq.~(\ref{eqn10}). 
The aspiration level is zero. The degrees of each agent are $8$ (black squares), $6$ (red circles), and $4$ (blue triangles).
Error bars are smaller than the symbol size.
}
\label{fig2}
\end{figure}

\begin{table}[b]
\caption{The values of fitting parameters $g_1$ and $g_2$ of Eq.~(\ref{eqn10}) in highly clustered RRGs with degree $k$.} 
\label{tab2}
\begin{tabular}{|c|l|l|l|l|l|}
\hline
$k$ & \multicolumn{1}{c|}{$4$} & \multicolumn{1}{c|}{$5$} & \multicolumn{1}{c|}{$6$} & \multicolumn{1}{c|}{$7$} & \multicolumn{1}{c|}{$8$} \\ \hline
$g_1$  & $0.251(4)$ & $0.259(6)$ & $0.247(6)$ & $0.254(6)$ & $0.248(1)$ \\ \hline
$g_2$ & $1.75(1)$ & $1.79(3)$ & $1.73(3)$ & $1.78(4)$ & $1.73(5)$ \\ \hline
\end{tabular}
\end{table}

We also studied critical behaviors in the highly clustered RRG, which are generated by the algorithm in Ref.~\onlinecite{Jeong_2019}. They are not always connected networks, but we verified that the proportion of the giant component is at least $99.9\%$. The degree of clustering is quantified by the clustering coefficient $c$ defined by 
\begin{eqnarray}
c = \frac{\sum_{i} c_i}{N}  ~\mbox{with } 
c_{i} = \left\{ \begin{array}{ll} \frac{2 l_i}{k_i (k_i-1)} & \mbox{for } k_i>1 \\ 0 & \mbox{for } k_i \leq 1
\end{array}
\right.,
\end{eqnarray}
where $k_i$ and $l_i$ are the degree and the number of links among the neighbors of agent $i$, respectively.\cite{watts1998collective} Figure~\ref{fig2} shows that critical selection noise ($\kappa_{c}$) decreases as the clustering coefficient ($c$) increases; we conjecture the relation
\begin{eqnarray}
\kappa_{c}(k,c) = \kappa_{c}(k,0) \frac{\ln(g_1 k + g_2-c)}{\ln(g_1 k + g_2)} . \label{eqn10}
\end{eqnarray}
Parameters $g_1$ and $g_2$ for a given $k$ are obtained by fitting and they are listed in Table~\ref{tab2}. Interestingly, the values of $g_1$ and $g_2$ are very close to $1/4$ and $7/4$, respectively. Notably, the decrease of the critical temperature by clustering in the Ising model is also observed in other small-world networks.\cite{Barrat,PhysRevE.64.057104} We found that the critical exponents do not depend on the clustering coefficient (see Table~\ref{tab1}). 

\subsection{\label{sec:level3-3}Effects of aspiration level on the phase transition of the coordination game}

\begin{figure}[t]
\centering
\includegraphics[angle=270,width=1\columnwidth]{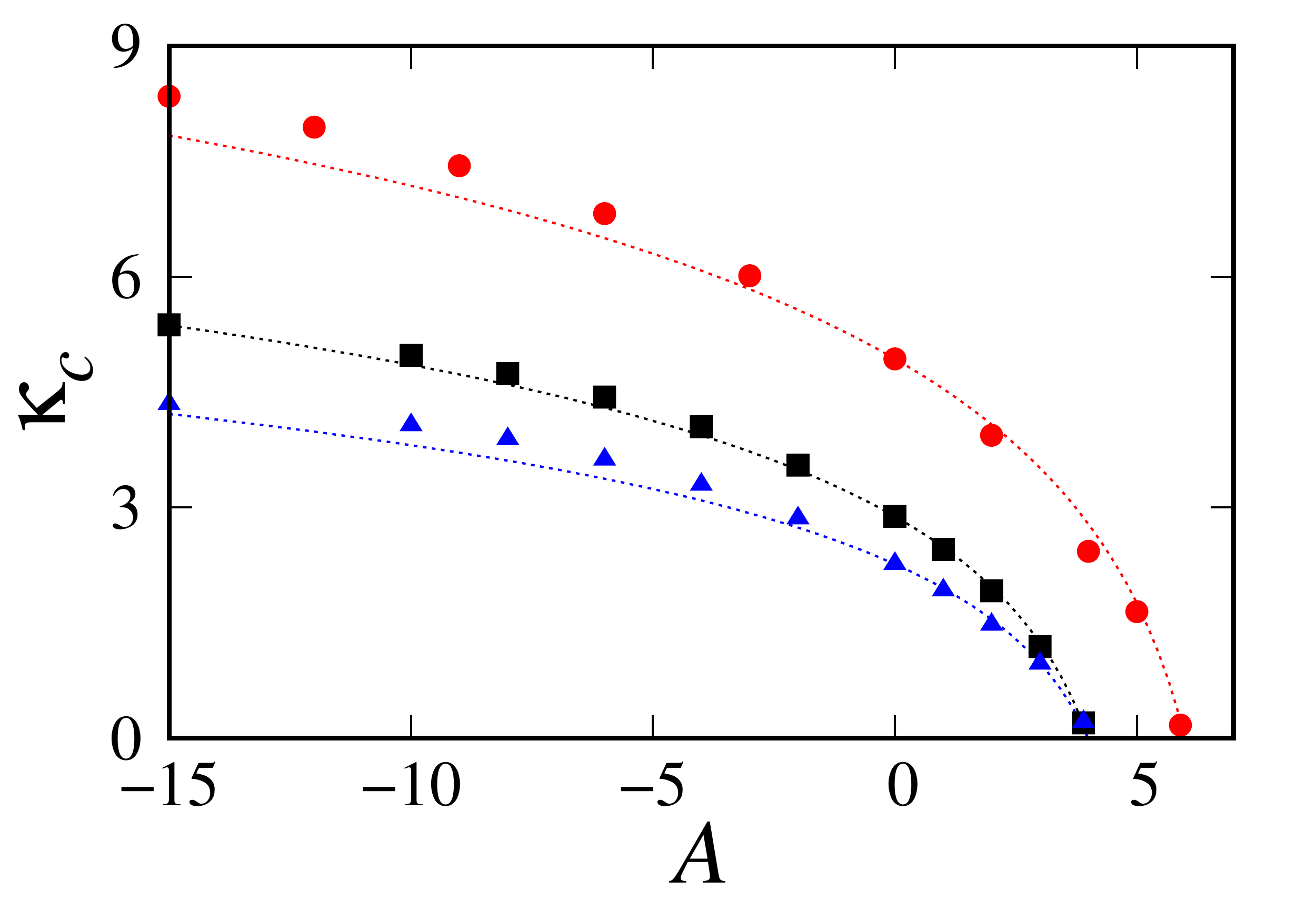}
\caption{The critical selection noise $\kappa_c$ as a function of aspiration level $A$ in the square lattice (blue triangles), RRG with $k=4$ (black squares), and RRG with $k=6$ (red circles).
Symbols represent the simulation results and dotted lines are from Eq.~(\ref{eqn12}).
Error bars are smaller than the symbol size.
}
\label{fig3}
\end{figure}

Now we present the results of the aspiration-based coordination game with non-zero aspiration level. 
The simulation methods are the same as the previous subsections. With varying the aspiration level $A$, we calculated critical selection noise $\kappa_{c}$ and critical exponents in the square lattice and RRGs without clustering. We found that $\kappa_{c}$ decreases as $A$ increases (see Fig.~\ref{fig3}), but the values of critical exponents are independent of aspiration level in the square lattice and RRGs (see Table~\ref{tab1}). We also confirmed the collapse of scaling functions regardless of the graph size for each value of $A$. Therefore, we conclude that the aspiration-based coordination game belongs to the same universality class as the Ising model in spite of nonequilibrium dynamics, as was conjectured by Ref.~\onlinecite{Grinstein}.

Critical selection noise $\kappa_{c}$ approaches zero at $A=k$ and the long-range order vanishes for higher $A$. Since there is a maximum value in the payoff an agent can obtain, the strategy update probability becomes too high to sustain the long-range order above a specific aspiration level $A$. We found no signature of long-range order at and above $A=k$. Interestingly, the critical aspiration level $A_c$ is the same as degree $k$ in both square lattice and RRG. Based on this result, we propose a trend function of
\begin{eqnarray}
\kappa_{c}(k,A) = \kappa_{c}(k,0) \frac{\ln(k+1-A)}{\ln(k+1)}. \label{eqn12}
\end{eqnarray}
As shown in Fig.~\ref{fig3}, Eq.~(\ref{eqn12}) fits well for positive $A$ but underestimates $\kappa_{c}(k,A)$ otherwise.

\subsection{\label{sec:level3-4} Aspiration-based coordination game in heterogeneous networks: hub centrality and local hubs}

\begin{table}[b]
\caption{Structural properties of the scale-free networks considered in this work: 
number of agents ($N$), average degree ($k_\mathrm{avg}$), clustering coefficient ($c$), and exponent of power-law degree distribution ($\delta$). BA network and HK network are generated by Barab\'{a}si-Albert model~\cite{Barabasi509} and Holme-Kim model,\cite{PhysRevE.65.026107} respectively. Airports network represents the connections of airports around the world.\cite{nr} Email network is email community network made public and posted by the Federal Energy Regulatory Commission.\cite{nr}}
\label{tab3}
\begin{tabular}{|l|l|l|l|l|}
\hline
\multicolumn{1}{|c|}{Networks} & \multicolumn{1}{c|}{$N$} & \multicolumn{1}{c|}{$k_\mathrm{avg}$} & \multicolumn{1}{c|}{$c$} & \multicolumn{1}{c|}{$\delta$} \\ \hline
BA network       & 3000  & $10$   & $\approx0$ & $3.0(1)$ \\ \hline
HK network       & 3000  & $10$   & $0.2281$   & $3.0(1)$ \\ \hline
Airports network & 2939  & $10.6$ & $0.4526$   & $1.7(2)$ \\ \hline
Email network    & 33696 & $10.7$ & $0.5091$   & $2.0(1)$ \\ \hline
\end{tabular}
\end{table}

\begin{figure}[t]
\centering
\includegraphics[angle=270,width=1\columnwidth]{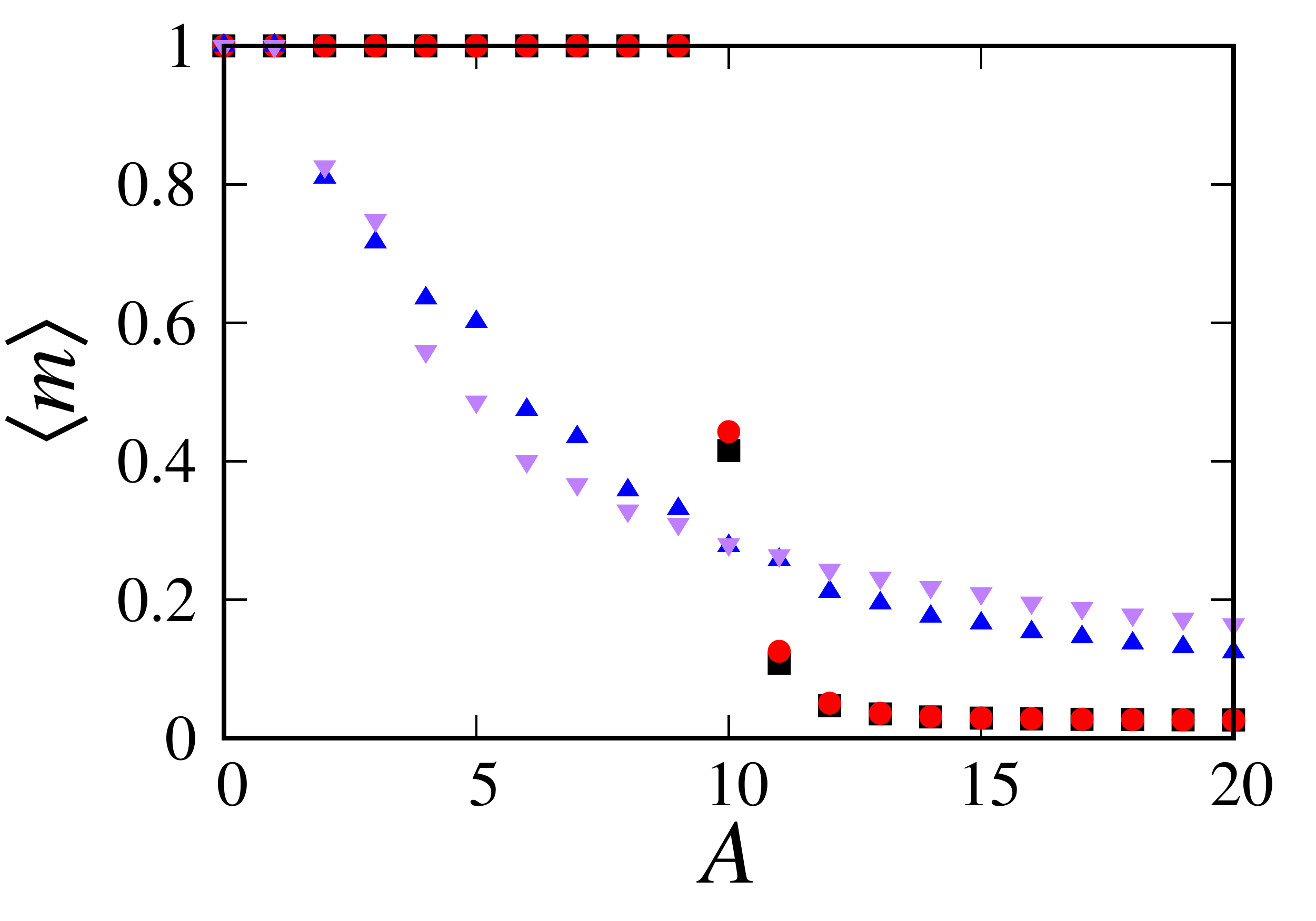}
\caption{Order parameter $\langle m \rangle$ as a function of $A$ at each network. 
The black squares are for BA scale-free network, the red circles are for the HK scale-free network, the blue triangles are for airports network, and the purple inverted triangles are for the email network. The selection noise is set to $\kappa=0.01$.
Error bars are smaller than the symbol size.}
\label{fig4}
\end{figure}

\begin{figure*}[t]
\centering
\includegraphics[angle=270,width=2\columnwidth]{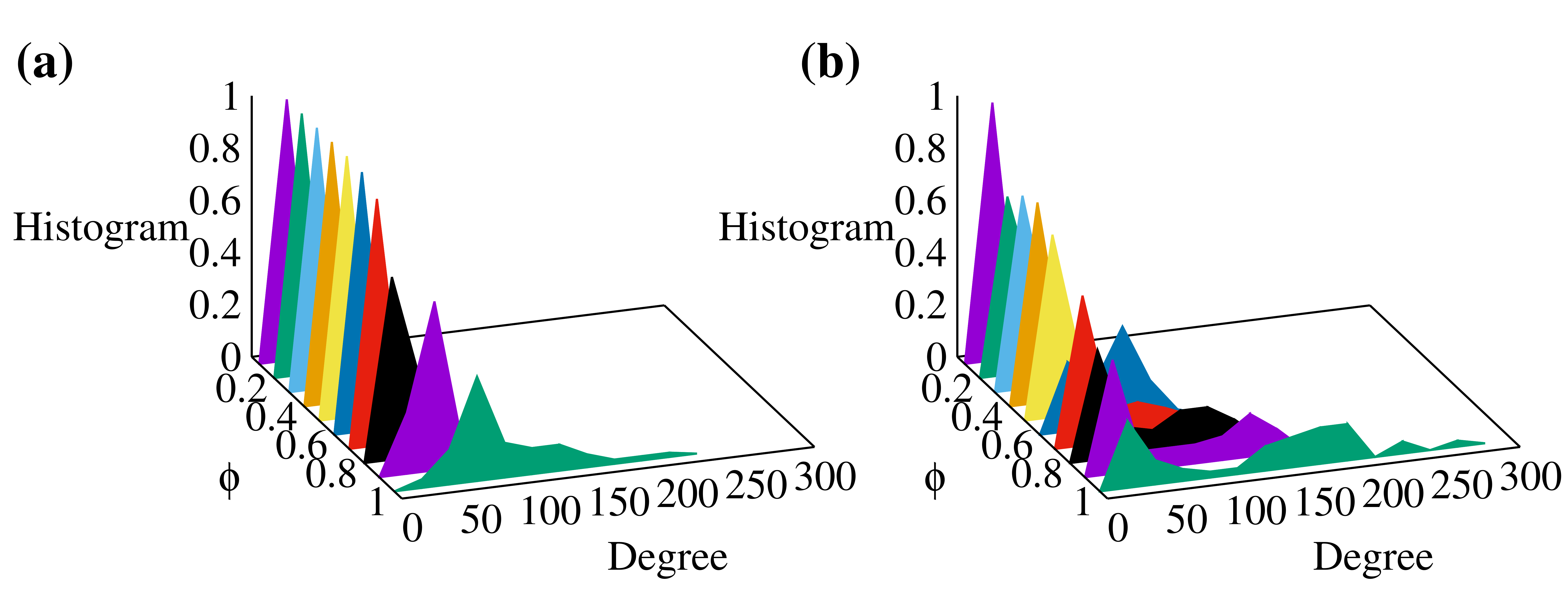}
\caption{
The histogram of the degree of agents at a given $\phi$ range. Hub centrality is divided at regular intervals of $0.1$ and the histogram bin of degree is $20$. (a) The histogram of BA network. (b) The histogram of airports network.
}
\label{fig5}
\end{figure*}

Up to here, we considered only regular graphs. In many cases, however, networks are highly heterogeneous~\cite{broido2019scale} and the heterogeneous degree distribution may affect the properties of the network dramatically.\cite{PhysRevE.69.026113,PhysRevLett.89.208701,Barabasi509,PhysRevLett.95.098104,PhysRevE.66.056118,WU20201} For example, the critical temperature of the Ising model is infinite in scale-free networks.\cite{BIANCONI2002166}

In this subsection, we study the effect of aspiration level at a low selection noise in two kinds of model networks and two real-world networks with scale-free degree distribution. Degree distributions of the four networks follow the power-law, $P(k) \propto k^{-\delta}$. Structural properties of the networks we consider in this subsection are presented in Table~\ref{tab3}. The exponent of power-law degree distribution $\delta$ in the BA and HK networks is~$3$ as expected. The simulations are performed at various aspiration levels. The selection noise is set to $\kappa=0.01$. Each simulation result was averaged over $10\,000$ generations after the system had reached an equilibrium state, and the final results were averaged over $30$ independent simulations. 

The results of $\langle m \rangle$ for the four kinds of networks are shown in Fig.~\ref{fig4}. For $A\leq0$, one of the two strategies dominates the networks, and $\langle m \rangle$ decreases to zero as $A$ increases for positive $A$. The change is very abrupt at $A=k_\mathrm{avg}$ in BA and HK networks. It is noteworthy that the BA and HK networks show almost the same behavior in spite of a large difference in clustering coefficient. In contrast, in the case of the real-world networks, $\langle m \rangle$ decreases gradually as $A$ increases. The real-world networks have lower $\langle m \rangle$ than the model networks for $A<k_\mathrm{avg}$ while it maintains a nonnegligible level for higher $A$.

To understand the remarkably different behaviors in model networks and real-world networks, we checked $\langle m \rangle$ of the agents with each degree near $A=k_\mathrm{avg}$, but we found no clue and we conclude that degree of an agent is not a dominating factor in coordination games in real-world networks. Therefore, we propose a new parameter hub centrality $\phi$, which determines the behavior of each agent in the coordination game beyond the degree. The value of hub centrality of an agent is defined as the proportion of the neighbors that have smaller degree than the agent among all its neighbors. For example, if an agent has five neighbors, three of which have neighbors less than five and the other two neighbors have degree of five or more, $\phi$ of the agent is $0.6$. A high value of $\phi$ means that the degree of the agent is high compared to those of its neighbors and the agent plays a leading role in its local environment. Since the hub centrality is determined solely by the degree of the agents in the network, it does not change in static networks.

We measured $\phi$ of all agents and Fig.~\ref{fig5} shows the histogram of degree for a given hub centrality range. Although the results of only BA scale-free network and airports network are shown, those of HK scale-free network and email network are qualitatively the same as Fig.~\ref{fig5}(a) and (b), respectively. In the case of model scale-free networks, the value of $\phi$ is proportional to a degree; the peak of histogram shifts to a higher value of the degree as hub centrality increases and all agents with degree more than $100$ have $\phi$ greater than $0.9$. Hub agents are connected with many low-degree agents and are affected critically by their strategies. In this network, sizable agents have a smaller degree than $k_\mathrm{avg}$ and cannot obtain enough payoff for $A\geq k_\mathrm{avg}$. They change their strategies so often to prevent their neighbors gain enough payoff. This causes the decrease of $\langle m \rangle$ and high-degree agents connected to them cannot keep their strategies, either.

\begin{figure}[b]
\centering
\includegraphics[angle=270,width=1\columnwidth]{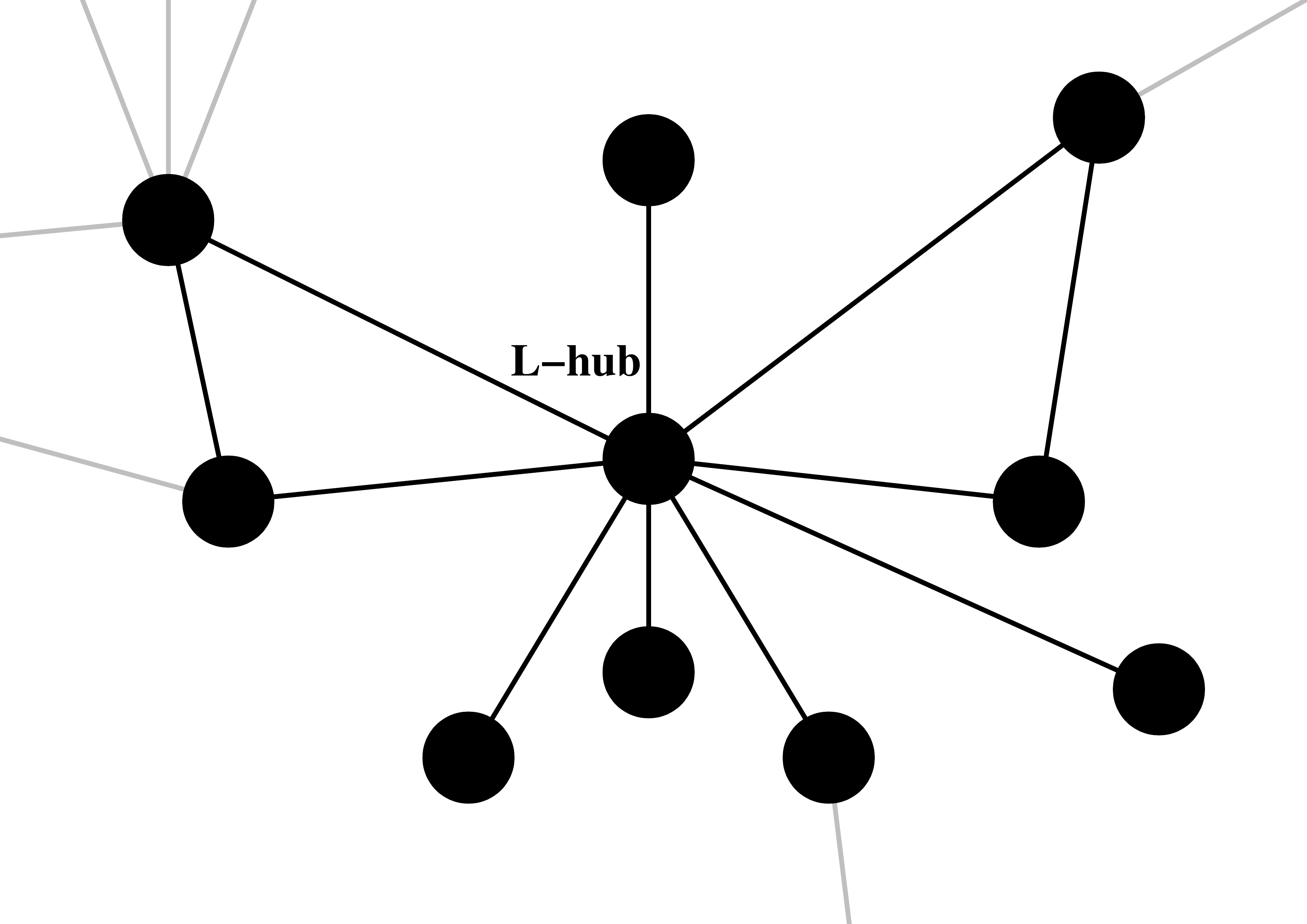}
\caption{A L-hub and its neighbors of airports network. The black solid lines represent connections within this subset. The gray solid lines represent connections between agents inside this subset and agents outside.}
\label{fig6}
\end{figure}

In real-world networks, to the contrary, the histogram separates into two parts as $\phi$ increases, as shown in Fig.~\ref{fig5}(b). In other words, there are a considerable number of agents that have a high $\phi$ and a low degree in real-world networks; some agents of $\phi>0.9$ have a degree lower than $k_\mathrm{avg}$. They are not hubs but behave like hubs locally, because a large portion of their neighbors has a lower degree. Therefore, we propose to call them local hubs (L-hubs). Figure~\ref{fig6} shows an example of an L-hub in the airports network; the degree of the agent is as low as $8$, but it is higher than all of its neighbors and the hub centrality of the agent is maximum ($\phi=1$). L-hubs have a relatively low degree and they are surrounded by neighbors of lower degree. Therefore, L-hubs and their neighbors can lose ordering, and $\langle m \rangle$ begins to decrease even for $A<k_\mathrm{avg}$. Moreover, agents with a high degree have various $\phi$ values: the agents who have a higher degree than $100$ have a broad range of $\phi$ from $0.6$ to $1.0$. This indicates that at least some agents have a considerable number of high-degree neighbors as well as many low-degree neighbors; the payoff of the agents is less influenced by the strategy of agents with low degrees. Thus, the agents can obtain enough payoff even in high aspiration level conditions to keep their ordering though agents with low degree change their strategies frequently.

\begin{figure*}
\centering
\includegraphics[angle=270,width=2\columnwidth]{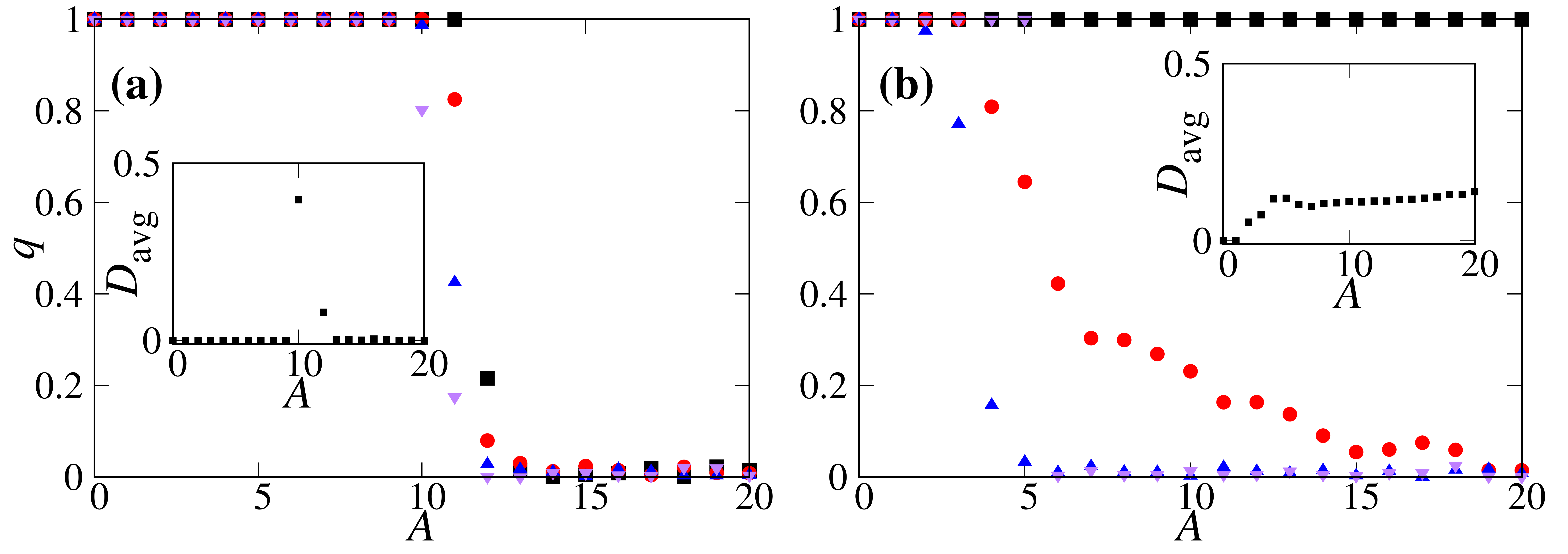}
\caption{Edwards-Anderson order parameter of four high-$\phi$ agents in BA scale-free network (the left panel) and in airports network (the right panel) as a function of $A$. The degrees of the agents are $131$ (black squares), $46$ (red circles), $35$ (blue triangles), and $16$ (purple inverted triangles) in (a); $172$ (black squares), $25$ (red circles), $9$ (blue triangles), and $6$ (purple inverted triangles) in (b). The agents represented by blue triangles and purple inverted triangles in (b) are classified as L-hubs. Insets: $D_\mathrm{avg}$ as a function of $A$ for each network, where $D_\mathrm{avg}$ is the average $D$ of high-$\phi$ agents ($\phi>0.9$).}
\label{fig7}
\end{figure*}

In order to examine the detailed behavior of each agent as $A$ changes, we measured the Edwards-Anderson order parameter ($q$),\cite{Edwards_1975,PhysRevB.72.024445,PhysRevE.91.062121} which is defined for agent $i$ as 
\begin{eqnarray}
q_i = \frac{1}{M} \left| \sum^{M}_{t} \alpha_i (t) \right| ,  \label{eqn13}
\end{eqnarray}
where $\alpha_i(t)=1$ if agent $i$ has the strategy S1 at generation $t$ and $\alpha_i(t)=-1$ otherwise. After the system reaches equilibrium, $\alpha_i$ is measured for each agent $M$ times, once per generation. 
We fixed $M$ to be $10\,000$. Large value of $q_i$ close to 1 means that the agent $i$ is frozen; if $q_i \approx 0$, the agent $i$ is melted. If the agent $i$ is partially frozen, $q_i$ is moderate value between $0$ and $1$.

In the melting process as $A$ increases, it is important to observe the change of Edwards-Anderson order parameter of agents with high $\phi$, which have considerably larger degrees than their neighbors and indicate the situation of freezing in their neighborhoods.

Figure~\ref{fig7} shows $q$ as a function of $A$ for four agents with high $\phi$ ($\phi>0.9$) in each of BA scale-free network and airports network. Two agents of airports network are classified as L-hub. There is no L-hub in BA scale-free network, where every agent with high $\phi$ has a higher degree than $k_\mathrm{avg}$. In the case of BA scale-free network, according to Fig.~\ref{fig7}(a), all the four agents are frozen for $A < k_\mathrm{avg}$ and they are melted abruptly for $A$ slightly larger than $k_\mathrm{avg}$. Agents with high $\phi$ in the HK scale-free network show the same behavior.

According to Fig.~\ref{fig7}(b), the two L-hubs (blue triangles and purple inverted triangles) are melted near $A = 6$. This means that the subset consisting of the L-hub and its neighbors is melted. Thus, $\langle m \rangle$ of the network begins to decrease at $A<k_\mathrm{avg}$.

Another agent, which is represented by red circles, shows a gradual decrease of $q$. When $A$ is zero, the agent is frozen. As $A$ increases, it becomes partially frozen in a specific range; some of its neighbors with a lower degree are melted and this leads to the increase of the strategy update probability of the agent. For $A>20$, the agent is melted completely. The last agent (black squares in Fig.~\ref{fig7}(b)), which has a very high degree, keeps its order up to $A=20$. In this case, although some of its neighbors are melted, the agent keeps its strategy interacting with a considerable number of neighbors with high degrees. Due to the presence of such agents, $\langle m \rangle$ is remarkably high at a high aspiration level. Email network also shows qualitatively the same behaviors.

To take a closer look at the melting process of high-$\phi$ agents ($\phi>0.9$) and their neighbors, we calculated the $q$-difference ($D$) of them, which is defined for agent $i$ as
\begin{eqnarray}
D_i = q_i - \frac{1}{k_i} \sum_{j \in \Omega_i} q_j , \label{eqn14}
\end{eqnarray}
where $\Omega_i$ is the set of neighbors of the agent $i$. If all of agent $i$ and its neighbors are melted or frozen, $D_i$ becomes almost zero. During the melting process, $D_i$ is expected to be positive since high-$\phi$ agents are melted at a larger $A$ than their neighbors. The average of the $q$-difference of high-$\phi$ agents ($D_\mathrm{avg}$) at each network is shown in the insets of Fig.~\ref{fig7}. 

As shown in the inset of Fig.~\ref{fig7}(a), $D_\mathrm{avg}$ has non-negligible values only in a very narrow region near $A=k_\mathrm{avg}$ in BA scale-free network. Immediately after some agents begin to melt near $A=k_\mathrm{avg}$, all agents lose the order abruptly as $A$ increases. In the case of airports network (inset of Fig.~\ref{fig7}(b)), to the contrary, $D_\mathrm{avg}$ has substantially high values in a wide range of $A$ even up to $A=20$. This implies that a considerable number of high-$\phi$ agents keep frozen or partially frozen states up to high $A$ though some of their neighbors are melted.

\section{\label{sec:level4}Conclusion}

We studied the coordination game with aspiration. We proved that this model coincides exactly with the Ising model in the zero aspiration level condition. We also confirmed this using the Monte-Carlo method in square lattice and RRGs. Particularly in the case of RRGs, we found that the critical selection noise decreases as the clustering coefficient increases.

With a non-zero aspiration level, the system also shows the phase transition but the critical selection noise decreases as the aspiration level increases. The phase transition and long-range order disappear when the aspiration level exceeds the degree of the regular graph. We showed that the critical exponents are independent of clustering and aspiration, and so the aspiration-based coordination game belongs to the same universality class as the Ising model at least for regular graphs.

Finally, we also carried out a study on the aspiration-based coordination game in four kinds of scale-free networks at a low selection noise. The order parameter decreases as the aspiration level increases in all the scale-free networks, but the behavior depends crucially on the network type: model network and real-world network. To understand the origin of the difference, we proposed the concept of the hub centrality and studied the relation between hub centrality and degree in the two kinds of networks. The hub centrality is proportional to the degree in the model networks, but the histogram of degree for high hub centrality separates into two parts in the real-world networks. Based on this observation, we concluded that there exist local hubs in real-world networks, which affect the coordination game dynamics seriously. We measured the Edwards-Anderson order parameter of agents of high hub centrality. It decreases abruptly near $A=k_{\mathrm{avg}}$ in model networks. In real-world networks, however, there are various cases depending on the local environment, and so the order parameter gradually decreases as the aspiration level increases.

Since the introduction of Barab\'{a}si-Albert scale-free network,\cite{Barabasi509} there have been several discussions about the differences between model networks and real-world networks.\cite{PhysRevLett.89.208701,broido2019scale,holme2019rare,PhysRevResearch.1.033034,PhysRevE.96.032306,Tsiotas6701} It was pointed out that the degree distribution of real-world networks is not exactly scale-free in most cases~\cite{broido2019scale,holme2019rare,PhysRevResearch.1.033034,PhysRevE.96.032306} and the assortativity is not negligible in some real-world networks.\cite{PhysRevLett.89.208701} However, it was not clear whether these differences may lead to critical differences in social network dynamics. 

In this work, we showed that model networks and real-world networks can show qualitatively different behaviors, which can not be understood by simple classic network parameters. We introduced new concepts of hub centrality and local hub to explain these behaviors. The histogram of hub centrality depends obviously on the network type and local hubs exist only in real-world networks. These differences result in the different behaviors of the order parameter and Edward-Anderson order parameter in the coordination game. Hub centrality shares some features with the concept of network hierarchy, and so it is expected to be more relevant in networks with hierarchical structures.\cite{clauset2008hierarchical} We are sure that these concepts would pave the way for deeper understandings of social networks and the dynamics in them.

\section*{Acknowledgments}
This work was supported by the National Research Foundation of Korea(NRF) grant funded by the Korea government(MSIT) (No. 2021R1F1A1052117).

\section*{DATA AVAILABILITY}
The data that support the findings of this study are available from the corresponding author upon reasonable request.

\section*{References}

\bibliography{References}% Produces the bibliography via BibTeX.

\begin{thebibliography}{70}%
\makeatletter
\providecommand \@ifxundefined [1]{%
 \@ifx{#1\undefined}
}%
\providecommand \@ifnum [1]{%
 \ifnum #1\expandafter \@firstoftwo
 \else \expandafter \@secondoftwo
 \fi
}%
\providecommand \@ifx [1]{%
 \ifx #1\expandafter \@firstoftwo
 \else \expandafter \@secondoftwo
 \fi
}%
\providecommand \natexlab [1]{#1}%
\providecommand \enquote  [1]{``#1''}%
\providecommand \bibnamefont  [1]{#1}%
\providecommand \bibfnamefont [1]{#1}%
\providecommand \citenamefont [1]{#1}%
\providecommand \href@noop [0]{\@secondoftwo}%
\providecommand \href [0]{\begingroup \@sanitize@url \@href}%
\providecommand \@href[1]{\@@startlink{#1}\@@href}%
\providecommand \@@href[1]{\endgroup#1\@@endlink}%
\providecommand \@sanitize@url [0]{\catcode `\\12\catcode `\$12\catcode
  `\&12\catcode `\#12\catcode `\^12\catcode `\_12\catcode `\%12\relax}%
\providecommand \@@startlink[1]{}%
\providecommand \@@endlink[0]{}%
\providecommand \url  [0]{\begingroup\@sanitize@url \@url }%
\providecommand \@url [1]{\endgroup\@href {#1}{\urlprefix }}%
\providecommand \urlprefix  [0]{URL }%
\providecommand \Eprint [0]{\href }%
\providecommand \doibase [0]{http://dx.doi.org/}%
\providecommand \selectlanguage [0]{\@gobble}%
\providecommand \bibinfo  [0]{\@secondoftwo}%
\providecommand \bibfield  [0]{\@secondoftwo}%
\providecommand \translation [1]{[#1]}%
\providecommand \BibitemOpen [0]{}%
\providecommand \bibitemStop [0]{}%
\providecommand \bibitemNoStop [0]{.\EOS\space}%
\providecommand \EOS [0]{\spacefactor3000\relax}%
\providecommand \BibitemShut  [1]{\csname bibitem#1\endcsname}%
\let\auto@bib@innerbib\@empty
%</preamble>
\bibitem [{\citenamefont {{Maynard}~Smith}\ and\ \citenamefont
  {Price}(1973)}]{smith1973}%
  \BibitemOpen
  \bibfield  {author} {\bibinfo {author} {\bibfnamefont {J.}~\bibnamefont
  {{Maynard}~Smith}}\ and\ \bibinfo {author} {\bibfnamefont {G.~R.}\
  \bibnamefont {Price}},\ }\bibfield  {title} {\enquote {\bibinfo {title} {The
  logic of animal conflict},}\ }\href {\doibase 10.1038/246015a0} {\bibfield
  {journal} {\bibinfo  {journal} {Nature}\ }\textbf {\bibinfo {volume} {246}},\
  \bibinfo {pages} {15--18} (\bibinfo {year} {1973})}\BibitemShut {NoStop}%
\bibitem [{\citenamefont {Nowak}(2006{\natexlab{a}})}]{nowak2006five}%
  \BibitemOpen
  \bibfield  {author} {\bibinfo {author} {\bibfnamefont {M.~A.}\ \bibnamefont
  {Nowak}},\ }\bibfield  {title} {\enquote {\bibinfo {title} {Five rules for
  the evolution of cooperation},}\ }\href {\doibase 10.1126/science.1133755}
  {\bibfield  {journal} {\bibinfo  {journal} {Science}\ }\textbf {\bibinfo
  {volume} {314}},\ \bibinfo {pages} {1560--1563} (\bibinfo {year}
  {2006}{\natexlab{a}})}\BibitemShut {NoStop}%
\bibitem [{\citenamefont {Castellano}, \citenamefont {Fortunato},\ and\
  \citenamefont {Loreto}(2009)}]{RevModPhys.81.591}%
  \BibitemOpen
  \bibfield  {author} {\bibinfo {author} {\bibfnamefont {C.}~\bibnamefont
  {Castellano}}, \bibinfo {author} {\bibfnamefont {S.}~\bibnamefont
  {Fortunato}}, \ and\ \bibinfo {author} {\bibfnamefont {V.}~\bibnamefont
  {Loreto}},\ }\bibfield  {title} {\enquote {\bibinfo {title} {Statistical
  physics of social dynamics},}\ }\href {\doibase 10.1103/RevModPhys.81.591}
  {\bibfield  {journal} {\bibinfo  {journal} {Rev. Mod. Phys.}\ }\textbf
  {\bibinfo {volume} {81}},\ \bibinfo {pages} {591--646} (\bibinfo {year}
  {2009})}\BibitemShut {NoStop}%
\bibitem [{\citenamefont {Fang}\ \emph {et~al.}(2019)\citenamefont {Fang},
  \citenamefont {Kruse}, \citenamefont {Lu},\ and\ \citenamefont
  {Wang}}]{RevModPhys.91.045004}%
  \BibitemOpen
  \bibfield  {author} {\bibinfo {author} {\bibfnamefont {X.}~\bibnamefont
  {Fang}}, \bibinfo {author} {\bibfnamefont {K.}~\bibnamefont {Kruse}},
  \bibinfo {author} {\bibfnamefont {T.}~\bibnamefont {Lu}}, \ and\ \bibinfo
  {author} {\bibfnamefont {J.}~\bibnamefont {Wang}},\ }\bibfield  {title}
  {\enquote {\bibinfo {title} {Nonequilibrium physics in biology},}\ }\href
  {\doibase 10.1103/RevModPhys.91.045004} {\bibfield  {journal} {\bibinfo
  {journal} {Rev. Mod. Phys.}\ }\textbf {\bibinfo {volume} {91}},\ \bibinfo
  {pages} {045004} (\bibinfo {year} {2019})}\BibitemShut {NoStop}%
\bibitem [{\citenamefont {Szab\'o}\ and\ \citenamefont
  {T\ifmmode~\mbox{\H{o}}\else \H{o}\fi{}ke}(1998)}]{PhysRevE.58.69}%
  \BibitemOpen
  \bibfield  {author} {\bibinfo {author} {\bibfnamefont {G.}~\bibnamefont
  {Szab\'o}}\ and\ \bibinfo {author} {\bibfnamefont {C.}~\bibnamefont
  {T\ifmmode~\mbox{\H{o}}\else \H{o}\fi{}ke}},\ }\bibfield  {title} {\enquote
  {\bibinfo {title} {Evolutionary prisoner's dilemma game on a square
  lattice},}\ }\href {\doibase 10.1103/PhysRevE.58.69} {\bibfield  {journal}
  {\bibinfo  {journal} {Phys. Rev. E}\ }\textbf {\bibinfo {volume} {58}},\
  \bibinfo {pages} {69--73} (\bibinfo {year} {1998})}\BibitemShut {NoStop}%
\bibitem [{\citenamefont {Nowak}\ and\ \citenamefont
  {May}(1992)}]{nowak1992evolutionary}%
  \BibitemOpen
  \bibfield  {author} {\bibinfo {author} {\bibfnamefont {M.~A.}\ \bibnamefont
  {Nowak}}\ and\ \bibinfo {author} {\bibfnamefont {R.~M.}\ \bibnamefont
  {May}},\ }\bibfield  {title} {\enquote {\bibinfo {title} {Evolutionary games
  and spatial chaos},}\ }\href {\doibase 10.1038/359826a0} {\bibfield
  {journal} {\bibinfo  {journal} {Nature}\ }\textbf {\bibinfo {volume} {359}},\
  \bibinfo {pages} {826--829} (\bibinfo {year} {1992})}\BibitemShut {NoStop}%
\bibitem [{\citenamefont {Pennisi}(2005)}]{Pennisi05}%
  \BibitemOpen
  \bibfield  {author} {\bibinfo {author} {\bibfnamefont {E.}~\bibnamefont
  {Pennisi}},\ }\bibfield  {title} {\enquote {\bibinfo {title} {How did
  cooperative behavior evolve{?}}}\ }\href {\doibase
  10.1126/science.309.5731.93} {\bibfield  {journal} {\bibinfo  {journal}
  {Science}\ }\textbf {\bibinfo {volume} {309}},\ \bibinfo {pages} {93--93}
  (\bibinfo {year} {2005})}\BibitemShut {NoStop}%
\bibitem [{\citenamefont {Tanimoto}(2015)}]{tanimoto2015fundamentals}%
  \BibitemOpen
  \bibfield  {author} {\bibinfo {author} {\bibfnamefont {J.}~\bibnamefont
  {Tanimoto}},\ }\href@noop {} {\emph {\bibinfo {title} {Fundamentals of
  evolutionary game theory and its applications}}}\ (\bibinfo  {publisher}
  {Springer},\ \bibinfo {address} {Tokyo},\ \bibinfo {year} {2015})\BibitemShut
  {NoStop}%
\bibitem [{\citenamefont {Pacheco}, \citenamefont {Traulsen},\ and\
  \citenamefont {Nowak}(2006)}]{PhysRevLett.97.258103}%
  \BibitemOpen
  \bibfield  {author} {\bibinfo {author} {\bibfnamefont {J.~M.}\ \bibnamefont
  {Pacheco}}, \bibinfo {author} {\bibfnamefont {A.}~\bibnamefont {Traulsen}}, \
  and\ \bibinfo {author} {\bibfnamefont {M.~A.}\ \bibnamefont {Nowak}},\
  }\bibfield  {title} {\enquote {\bibinfo {title} {Coevolution of strategy and
  structure in complex networks with dynamical linking},}\ }\href {\doibase
  10.1103/PhysRevLett.97.258103} {\bibfield  {journal} {\bibinfo  {journal}
  {Phys. Rev. Lett.}\ }\textbf {\bibinfo {volume} {97}},\ \bibinfo {pages}
  {258103} (\bibinfo {year} {2006})}\BibitemShut {NoStop}%
\bibitem [{\citenamefont {Fu}\ \emph {et~al.}(2021)\citenamefont {Fu},
  \citenamefont {Zhang}, \citenamefont {Guo},\ and\ \citenamefont
  {Xie}}]{Fu2021evolution}%
  \BibitemOpen
  \bibfield  {author} {\bibinfo {author} {\bibfnamefont {Y.}~\bibnamefont
  {Fu}}, \bibinfo {author} {\bibfnamefont {Y.}~\bibnamefont {Zhang}}, \bibinfo
  {author} {\bibfnamefont {Y.}~\bibnamefont {Guo}}, \ and\ \bibinfo {author}
  {\bibfnamefont {Y.}~\bibnamefont {Xie}},\ }\bibfield  {title} {\enquote
  {\bibinfo {title} {Evolutionary dynamics of cooperation with the celebrity
  effect in complex networks},}\ }\href {\doibase 10.1063/5.0033335} {\bibfield
   {journal} {\bibinfo  {journal} {Chaos}\ }\textbf {\bibinfo {volume} {31}},\
  \bibinfo {pages} {013130} (\bibinfo {year} {2021})}\BibitemShut {NoStop}%
\bibitem [{\citenamefont {Xu}\ \emph {et~al.}(2021)\citenamefont {Xu},
  \citenamefont {Wang}, \citenamefont {Han}, \citenamefont {He},\ and\
  \citenamefont {Wang}}]{XU2021110380}%
  \BibitemOpen
  \bibfield  {author} {\bibinfo {author} {\bibfnamefont {B.}~\bibnamefont
  {Xu}}, \bibinfo {author} {\bibfnamefont {Y.}~\bibnamefont {Wang}}, \bibinfo
  {author} {\bibfnamefont {Y.}~\bibnamefont {Han}}, \bibinfo {author}
  {\bibfnamefont {Y.}~\bibnamefont {He}}, \ and\ \bibinfo {author}
  {\bibfnamefont {Z.}~\bibnamefont {Wang}},\ }\bibfield  {title} {\enquote
  {\bibinfo {title} {Interaction patterns and coordination in two population
  groups: A dynamic perspective},}\ }\href {\doibase
  10.1016/j.chaos.2020.110380} {\bibfield  {journal} {\bibinfo  {journal}
  {Chaos, Soliton. Fract.}\ }\textbf {\bibinfo {volume} {142}},\ \bibinfo
  {pages} {110380} (\bibinfo {year} {2021})}\BibitemShut {NoStop}%
\bibitem [{\citenamefont {Weibull}(1995)}]{weibull1995evolutionary}%
  \BibitemOpen
  \bibfield  {author} {\bibinfo {author} {\bibfnamefont {J.~W.}\ \bibnamefont
  {Weibull}},\ }\href@noop {} {\emph {\bibinfo {title} {Evolutionary game
  theory}}}\ (\bibinfo  {publisher} {MIT Press},\ \bibinfo {address}
  {Cambridge, MA},\ \bibinfo {year} {1995})\BibitemShut {NoStop}%
\bibitem [{\citenamefont {Szab\'o}\ and\ \citenamefont
  {Kir\'aly}(2016)}]{PhysRevE.93.052108}%
  \BibitemOpen
  \bibfield  {author} {\bibinfo {author} {\bibfnamefont {G.}~\bibnamefont
  {Szab\'o}}\ and\ \bibinfo {author} {\bibfnamefont {B.}~\bibnamefont
  {Kir\'aly}},\ }\bibfield  {title} {\enquote {\bibinfo {title} {Extension of a
  spatial evolutionary coordination game with neutral options},}\ }\href
  {\doibase 10.1103/PhysRevE.93.052108} {\bibfield  {journal} {\bibinfo
  {journal} {Phys. Rev. E}\ }\textbf {\bibinfo {volume} {93}},\ \bibinfo
  {pages} {052108} (\bibinfo {year} {2016})}\BibitemShut {NoStop}%
\bibitem [{\citenamefont {Kir\'aly}\ and\ \citenamefont
  {Szab\'o}(2017{\natexlab{a}})}]{PhysRevE.95.012303}%
  \BibitemOpen
  \bibfield  {author} {\bibinfo {author} {\bibfnamefont {B.}~\bibnamefont
  {Kir\'aly}}\ and\ \bibinfo {author} {\bibfnamefont {G.}~\bibnamefont
  {Szab\'o}},\ }\bibfield  {title} {\enquote {\bibinfo {title} {Evolutionary
  games with coordination and self-dependent interactions},}\ }\href {\doibase
  10.1103/PhysRevE.95.012303} {\bibfield  {journal} {\bibinfo  {journal} {Phys.
  Rev. E}\ }\textbf {\bibinfo {volume} {95}},\ \bibinfo {pages} {012303}
  (\bibinfo {year} {2017}{\natexlab{a}})}\BibitemShut {NoStop}%
\bibitem [{\citenamefont {Kir\'aly}\ and\ \citenamefont
  {Szab\'o}(2017{\natexlab{b}})}]{PhysRevE.96.042101}%
  \BibitemOpen
  \bibfield  {author} {\bibinfo {author} {\bibfnamefont {B.}~\bibnamefont
  {Kir\'aly}}\ and\ \bibinfo {author} {\bibfnamefont {G.}~\bibnamefont
  {Szab\'o}},\ }\bibfield  {title} {\enquote {\bibinfo {title} {Evolutionary
  games combining two or three pair coordinations on a square lattice},}\
  }\href {\doibase 10.1103/PhysRevE.96.042101} {\bibfield  {journal} {\bibinfo
  {journal} {Phys. Rev. E}\ }\textbf {\bibinfo {volume} {96}},\ \bibinfo
  {pages} {042101} (\bibinfo {year} {2017}{\natexlab{b}})}\BibitemShut
  {NoStop}%
\bibitem [{\citenamefont {Jin}\ and\ \citenamefont {Yu}(2021)}]{Jin21}%
  \BibitemOpen
  \bibfield  {author} {\bibinfo {author} {\bibfnamefont {K.}~\bibnamefont
  {Jin}}\ and\ \bibinfo {author} {\bibfnamefont {U.}~\bibnamefont {Yu}},\
  }\bibfield  {title} {\enquote {\bibinfo {title} {Reference to global state
  and social contagion dynamics},}\ }\href {\doibase 10.3389/fphy.2021.684223}
  {\bibfield  {journal} {\bibinfo  {journal} {Front. Phys.}\ }\textbf {\bibinfo
  {volume} {9}},\ \bibinfo {pages} {254} (\bibinfo {year} {2021})}\BibitemShut
  {NoStop}%
\bibitem [{\citenamefont {Nowak}(2006{\natexlab{b}})}]{nowak2006evolutionary}%
  \BibitemOpen
  \bibfield  {author} {\bibinfo {author} {\bibfnamefont {M.~A.}\ \bibnamefont
  {Nowak}},\ }\href@noop {} {\emph {\bibinfo {title} {Evolutionary dynamics:
  exploring the equations of life}}}\ (\bibinfo  {publisher} {Harvard
  university press},\ \bibinfo {address} {Cambridge, Massachusetts},\ \bibinfo
  {year} {2006})\BibitemShut {NoStop}%
\bibitem [{\citenamefont {Wu}\ and\ \citenamefont
  {Holme}(2009)}]{PhysRevE.80.026108}%
  \BibitemOpen
  \bibfield  {author} {\bibinfo {author} {\bibfnamefont {Z.-X.}\ \bibnamefont
  {Wu}}\ and\ \bibinfo {author} {\bibfnamefont {P.}~\bibnamefont {Holme}},\
  }\bibfield  {title} {\enquote {\bibinfo {title} {Effects of
  strategy-migration direction and noise in the evolutionary spatial
  prisoner’s dilemma},}\ }\href {\doibase 10.1103/PhysRevE.80.026108}
  {\bibfield  {journal} {\bibinfo  {journal} {Phys. Rev. E}\ }\textbf {\bibinfo
  {volume} {80}},\ \bibinfo {pages} {026108} (\bibinfo {year}
  {2009})}\BibitemShut {NoStop}%
\bibitem [{\citenamefont {Amaral}\ \emph {et~al.}(2016)\citenamefont {Amaral},
  \citenamefont {Wardil}, \citenamefont {Perc},\ and\ \citenamefont
  {da~Silva}}]{PhysRevE.94.032317}%
  \BibitemOpen
  \bibfield  {author} {\bibinfo {author} {\bibfnamefont {M.~A.}\ \bibnamefont
  {Amaral}}, \bibinfo {author} {\bibfnamefont {L.}~\bibnamefont {Wardil}},
  \bibinfo {author} {\bibfnamefont {M.}~\bibnamefont {Perc}}, \ and\ \bibinfo
  {author} {\bibfnamefont {J.~K.~L.}\ \bibnamefont {da~Silva}},\ }\bibfield
  {title} {\enquote {\bibinfo {title} {Stochastic win-stay-lose-shift strategy
  with dynamic aspirations in evolutionary social dilemmas},}\ }\href {\doibase
  10.1103/PhysRevE.94.032317} {\bibfield  {journal} {\bibinfo  {journal} {Phys.
  Rev. E}\ }\textbf {\bibinfo {volume} {94}},\ \bibinfo {pages} {032317}
  (\bibinfo {year} {2016})}\BibitemShut {NoStop}%
\bibitem [{\citenamefont {Chen}\ and\ \citenamefont
  {Wang}(2008)}]{PhysRevE.77.017103}%
  \BibitemOpen
  \bibfield  {author} {\bibinfo {author} {\bibfnamefont {X.}~\bibnamefont
  {Chen}}\ and\ \bibinfo {author} {\bibfnamefont {L.}~\bibnamefont {Wang}},\
  }\bibfield  {title} {\enquote {\bibinfo {title} {Promotion of cooperation
  induced by appropriate payoff aspirations in a small-world networked game},}\
  }\href {\doibase 10.1103/PhysRevE.77.017103} {\bibfield  {journal} {\bibinfo
  {journal} {Phys. Rev. E}\ }\textbf {\bibinfo {volume} {77}},\ \bibinfo
  {pages} {017103} (\bibinfo {year} {2008})}\BibitemShut {NoStop}%
\bibitem [{\citenamefont {Amaral}\ and\ \citenamefont
  {Javarone}(2020)}]{PhysRevE.101.062309}%
  \BibitemOpen
  \bibfield  {author} {\bibinfo {author} {\bibfnamefont {M.~A.}\ \bibnamefont
  {Amaral}}\ and\ \bibinfo {author} {\bibfnamefont {M.~A.}\ \bibnamefont
  {Javarone}},\ }\bibfield  {title} {\enquote {\bibinfo {title} {Strategy
  equilibrium in dilemma games with off-diagonal payoff perturbations},}\
  }\href {\doibase 10.1103/PhysRevE.101.062309} {\bibfield  {journal} {\bibinfo
   {journal} {Phys. Rev. E}\ }\textbf {\bibinfo {volume} {101}},\ \bibinfo
  {pages} {062309} (\bibinfo {year} {2020})}\BibitemShut {NoStop}%
\bibitem [{\citenamefont {Szab\'o}\ and\ \citenamefont
  {F\'{a}th}(2007)}]{SZABO200797}%
  \BibitemOpen
  \bibfield  {author} {\bibinfo {author} {\bibfnamefont {G.}~\bibnamefont
  {Szab\'o}}\ and\ \bibinfo {author} {\bibfnamefont {G.}~\bibnamefont
  {F\'{a}th}},\ }\bibfield  {title} {\enquote {\bibinfo {title} {Evolutionary
  games on graphs},}\ }\href {\doibase 10.1016/j.physrep.2007.04.004}
  {\bibfield  {journal} {\bibinfo  {journal} {Phys. Rep.}\ }\textbf {\bibinfo
  {volume} {446}},\ \bibinfo {pages} {97--216} (\bibinfo {year}
  {2007})}\BibitemShut {NoStop}%
\bibitem [{\citenamefont {Nowak}\ and\ \citenamefont
  {Sigmund}(1993)}]{Nowak93}%
  \BibitemOpen
  \bibfield  {author} {\bibinfo {author} {\bibfnamefont {M.}~\bibnamefont
  {Nowak}}\ and\ \bibinfo {author} {\bibfnamefont {K.}~\bibnamefont
  {Sigmund}},\ }\bibfield  {title} {\enquote {\bibinfo {title} {A strategy of
  win-stay, lose-shift that outperforms tit-for-tat in the prisoner's dilemma
  game},}\ }\href {\doibase 10.1038/364056a0} {\bibfield  {journal} {\bibinfo
  {journal} {Nature}\ }\textbf {\bibinfo {volume} {364}},\ \bibinfo {pages}
  {56--58} (\bibinfo {year} {1993})}\BibitemShut {NoStop}%
\bibitem [{\citenamefont {Liu}\ \emph {et~al.}(2011)\citenamefont {Liu},
  \citenamefont {Chen}, \citenamefont {Wang}, \citenamefont {Li}, \citenamefont
  {Zhang},\ and\ \citenamefont {Wang}}]{Liu11}%
  \BibitemOpen
  \bibfield  {author} {\bibinfo {author} {\bibfnamefont {Y.}~\bibnamefont
  {Liu}}, \bibinfo {author} {\bibfnamefont {X.}~\bibnamefont {Chen}}, \bibinfo
  {author} {\bibfnamefont {L.}~\bibnamefont {Wang}}, \bibinfo {author}
  {\bibfnamefont {B.}~\bibnamefont {Li}}, \bibinfo {author} {\bibfnamefont
  {W.}~\bibnamefont {Zhang}}, \ and\ \bibinfo {author} {\bibfnamefont
  {H.}~\bibnamefont {Wang}},\ }\bibfield  {title} {\enquote {\bibinfo {title}
  {Aspiration-based learning promotes cooperation in spatial prisoner's dilemma
  games},}\ }\href {\doibase 10.1209/0295-5075/94/60002} {\bibfield  {journal}
  {\bibinfo  {journal} {{EPL} (Europhys. Lett.)}\ }\textbf {\bibinfo {volume}
  {94}},\ \bibinfo {pages} {60002} (\bibinfo {year} {2011})}\BibitemShut
  {NoStop}%
\bibitem [{\citenamefont {Santos}\ and\ \citenamefont
  {Pacheco}(2005)}]{PhysRevLett.95.098104}%
  \BibitemOpen
  \bibfield  {author} {\bibinfo {author} {\bibfnamefont {F.~C.}\ \bibnamefont
  {Santos}}\ and\ \bibinfo {author} {\bibfnamefont {J.~M.}\ \bibnamefont
  {Pacheco}},\ }\bibfield  {title} {\enquote {\bibinfo {title} {Scale-free
  networks provide a unifying framework for the emergence of cooperation},}\
  }\href {\doibase 10.1103/PhysRevLett.95.098104} {\bibfield  {journal}
  {\bibinfo  {journal} {Phys. Rev. Lett.}\ }\textbf {\bibinfo {volume} {95}},\
  \bibinfo {pages} {098104} (\bibinfo {year} {2005})}\BibitemShut {NoStop}%
\bibitem [{\citenamefont {Holme}\ \emph {et~al.}(2003)\citenamefont {Holme},
  \citenamefont {Trusina}, \citenamefont {Kim},\ and\ \citenamefont
  {Minnhagen}}]{PhysRevE.68.030901}%
  \BibitemOpen
  \bibfield  {author} {\bibinfo {author} {\bibfnamefont {P.}~\bibnamefont
  {Holme}}, \bibinfo {author} {\bibfnamefont {A.}~\bibnamefont {Trusina}},
  \bibinfo {author} {\bibfnamefont {B.~J.}\ \bibnamefont {Kim}}, \ and\
  \bibinfo {author} {\bibfnamefont {P.}~\bibnamefont {Minnhagen}},\ }\bibfield
  {title} {\enquote {\bibinfo {title} {Prisoners’ dilemma in real-world
  acquaintance networks: Spikes and quasiequilibria induced by the interplay
  between structure and dynamics},}\ }\href {\doibase
  10.1103/PhysRevE.68.030901} {\bibfield  {journal} {\bibinfo  {journal} {Phys.
  Rev. E}\ }\textbf {\bibinfo {volume} {68}},\ \bibinfo {pages} {030901}
  (\bibinfo {year} {2003})}\BibitemShut {NoStop}%
\bibitem [{\citenamefont {McAvoy}, \citenamefont {Allen},\ and\ \citenamefont
  {Nowak}(2020)}]{mcavoy2020social}%
  \BibitemOpen
  \bibfield  {author} {\bibinfo {author} {\bibfnamefont {A.}~\bibnamefont
  {McAvoy}}, \bibinfo {author} {\bibfnamefont {B.}~\bibnamefont {Allen}}, \
  and\ \bibinfo {author} {\bibfnamefont {M.~A.}\ \bibnamefont {Nowak}},\
  }\bibfield  {title} {\enquote {\bibinfo {title} {Social goods dilemmas in
  heterogeneous societies},}\ }\href {\doibase 10.1038/s41562-020-0881-2}
  {\bibfield  {journal} {\bibinfo  {journal} {Nat. Hum. Behav.}\ }\textbf
  {\bibinfo {volume} {4}},\ \bibinfo {pages} {819--831} (\bibinfo {year}
  {2020})}\BibitemShut {NoStop}%
\bibitem [{\citenamefont {Alvarez-Rodriguez}\ \emph {et~al.}(2021)\citenamefont
  {Alvarez-Rodriguez}, \citenamefont {Battiston}, \citenamefont {de~Arruda},
  \citenamefont {Moreno}, \citenamefont {Perc},\ and\ \citenamefont
  {Latora}}]{alvarez2021evolutionary}%
  \BibitemOpen
  \bibfield  {author} {\bibinfo {author} {\bibfnamefont {U.}~\bibnamefont
  {Alvarez-Rodriguez}}, \bibinfo {author} {\bibfnamefont {F.}~\bibnamefont
  {Battiston}}, \bibinfo {author} {\bibfnamefont {G.~F.}\ \bibnamefont
  {de~Arruda}}, \bibinfo {author} {\bibfnamefont {Y.}~\bibnamefont {Moreno}},
  \bibinfo {author} {\bibfnamefont {M.}~\bibnamefont {Perc}}, \ and\ \bibinfo
  {author} {\bibfnamefont {V.}~\bibnamefont {Latora}},\ }\bibfield  {title}
  {\enquote {\bibinfo {title} {Evolutionary dynamics of higher-order
  interactions in social networks},}\ }\href {\doibase
  10.1038/s41562-020-01024-1} {\bibfield  {journal} {\bibinfo  {journal} {Nat.
  Hum. Behav.}\ }\textbf {\bibinfo {volume} {1}},\ \bibinfo {pages} {1--10}
  (\bibinfo {year} {2021})}\BibitemShut {NoStop}%
\bibitem [{\citenamefont {Santos}, \citenamefont {Pacheco},\ and\ \citenamefont
  {Lenaerts}(2006)}]{Santos3490}%
  \BibitemOpen
  \bibfield  {author} {\bibinfo {author} {\bibfnamefont {F.~C.}\ \bibnamefont
  {Santos}}, \bibinfo {author} {\bibfnamefont {J.~M.}\ \bibnamefont {Pacheco}},
  \ and\ \bibinfo {author} {\bibfnamefont {T.}~\bibnamefont {Lenaerts}},\
  }\bibfield  {title} {\enquote {\bibinfo {title} {Evolutionary dynamics of
  social dilemmas in structured heterogeneous populations},}\ }\href {\doibase
  10.1073/pnas.0508201103} {\bibfield  {journal} {\bibinfo  {journal} {Proc.
  Natl. Acad. Sci. U.S.A.}\ }\textbf {\bibinfo {volume} {103}},\ \bibinfo
  {pages} {3490--3494} (\bibinfo {year} {2006})}\BibitemShut {NoStop}%
\bibitem [{\citenamefont {Watts}\ and\ \citenamefont
  {Strogatz}(1998)}]{watts1998collective}%
  \BibitemOpen
  \bibfield  {author} {\bibinfo {author} {\bibfnamefont {D.~J.}\ \bibnamefont
  {Watts}}\ and\ \bibinfo {author} {\bibfnamefont {S.~H.}\ \bibnamefont
  {Strogatz}},\ }\bibfield  {title} {\enquote {\bibinfo {title} {Collective
  dynamics of ‘small-world’ networks},}\ }\href {\doibase 10.1038/30918}
  {\bibfield  {journal} {\bibinfo  {journal} {Nature}\ }\textbf {\bibinfo
  {volume} {393}},\ \bibinfo {pages} {440--442} (\bibinfo {year}
  {1998})}\BibitemShut {NoStop}%
\bibitem [{\citenamefont {Barab{\'a}si}\ and\ \citenamefont
  {Albert}(1999)}]{Barabasi509}%
  \BibitemOpen
  \bibfield  {author} {\bibinfo {author} {\bibfnamefont {A.-L.}\ \bibnamefont
  {Barab{\'a}si}}\ and\ \bibinfo {author} {\bibfnamefont {R.}~\bibnamefont
  {Albert}},\ }\bibfield  {title} {\enquote {\bibinfo {title} {Emergence of
  scaling in random networks},}\ }\href {\doibase 10.1126/science.286.5439.509}
  {\bibfield  {journal} {\bibinfo  {journal} {Science}\ }\textbf {\bibinfo
  {volume} {286}},\ \bibinfo {pages} {509--512} (\bibinfo {year}
  {1999})}\BibitemShut {NoStop}%
\bibitem [{\citenamefont {Newman}(2002)}]{PhysRevLett.89.208701}%
  \BibitemOpen
  \bibfield  {author} {\bibinfo {author} {\bibfnamefont {M.~E.~J.}\
  \bibnamefont {Newman}},\ }\bibfield  {title} {\enquote {\bibinfo {title}
  {Assortative mixing in networks},}\ }\href {\doibase
  10.1103/PhysRevLett.89.208701} {\bibfield  {journal} {\bibinfo  {journal}
  {Phys. Rev. Lett.}\ }\textbf {\bibinfo {volume} {89}},\ \bibinfo {pages}
  {208701} (\bibinfo {year} {2002})}\BibitemShut {NoStop}%
\bibitem [{\citenamefont {Holme}\ and\ \citenamefont
  {Kim}(2002)}]{PhysRevE.65.026107}%
  \BibitemOpen
  \bibfield  {author} {\bibinfo {author} {\bibfnamefont {P.}~\bibnamefont
  {Holme}}\ and\ \bibinfo {author} {\bibfnamefont {B.~J.}\ \bibnamefont
  {Kim}},\ }\bibfield  {title} {\enquote {\bibinfo {title} {Growing scale-free
  networks with tunable clustering},}\ }\href {\doibase
  10.1103/PhysRevE.65.026107} {\bibfield  {journal} {\bibinfo  {journal} {Phys.
  Rev. E}\ }\textbf {\bibinfo {volume} {65}},\ \bibinfo {pages} {026107}
  (\bibinfo {year} {2002})}\BibitemShut {NoStop}%
\bibitem [{\citenamefont {Bollob\'{a}s}(2001)}]{Bollobas01}%
  \BibitemOpen
  \bibfield  {author} {\bibinfo {author} {\bibfnamefont {B.}~\bibnamefont
  {Bollob\'{a}s}},\ }\href {\doibase 10.1017/CBO9780511814068} {\emph {\bibinfo
  {title} {Random Graphs}}},\ \bibinfo {edition} {2nd}\ ed.,\ Cambridge Studies
  in Advanced Mathematics\ (\bibinfo  {publisher} {Cambridge University
  Press},\ \bibinfo {address} {Cambridge},\ \bibinfo {year} {2001})\BibitemShut
  {NoStop}%
\bibitem [{\citenamefont {Jeong}, \citenamefont {Jang},\ and\ \citenamefont
  {Yu}(2019)}]{Jeong_2019}%
  \BibitemOpen
  \bibfield  {author} {\bibinfo {author} {\bibfnamefont {W.}~\bibnamefont
  {Jeong}}, \bibinfo {author} {\bibfnamefont {H.}~\bibnamefont {Jang}}, \ and\
  \bibinfo {author} {\bibfnamefont {U.}~\bibnamefont {Yu}},\ }\bibfield
  {title} {\enquote {\bibinfo {title} {Highly clustered complex networks in the
  configuration model: Random regular small-world network},}\ }\href {\doibase
  10.1209/0295-5075/128/16001} {\bibfield  {journal} {\bibinfo  {journal}
  {{EPL} (Europhys. Lett.)}\ }\textbf {\bibinfo {volume} {128}},\ \bibinfo
  {pages} {16001} (\bibinfo {year} {2019})}\BibitemShut {NoStop}%
\bibitem [{\citenamefont {Broido}\ and\ \citenamefont
  {Clauset}(2019)}]{broido2019scale}%
  \BibitemOpen
  \bibfield  {author} {\bibinfo {author} {\bibfnamefont {A.~D.}\ \bibnamefont
  {Broido}}\ and\ \bibinfo {author} {\bibfnamefont {A.}~\bibnamefont
  {Clauset}},\ }\bibfield  {title} {\enquote {\bibinfo {title} {Scale-free
  networks are rare},}\ }\href {\doibase 10.1038/s41467-019-08746-5} {\bibfield
   {journal} {\bibinfo  {journal} {Nat. Commun.}\ }\textbf {\bibinfo {volume}
  {10}},\ \bibinfo {pages} {1--10} (\bibinfo {year} {2019})}\BibitemShut
  {NoStop}%
\bibitem [{\citenamefont {Holme}(2019)}]{holme2019rare}%
  \BibitemOpen
  \bibfield  {author} {\bibinfo {author} {\bibfnamefont {P.}~\bibnamefont
  {Holme}},\ }\bibfield  {title} {\enquote {\bibinfo {title} {Rare and
  everywhere: Perspectives on scale-free networks},}\ }\href {\doibase
  10.1038/s41467-019-09038-8} {\bibfield  {journal} {\bibinfo  {journal} {Nat.
  Commun.}\ }\textbf {\bibinfo {volume} {10}},\ \bibinfo {pages} {1--3}
  (\bibinfo {year} {2019})}\BibitemShut {NoStop}%
\bibitem [{\citenamefont {Chen}, \citenamefont {Lin},\ and\ \citenamefont
  {Wu}(2007)}]{CHEN2007379}%
  \BibitemOpen
  \bibfield  {author} {\bibinfo {author} {\bibfnamefont {Y.-S.}\ \bibnamefont
  {Chen}}, \bibinfo {author} {\bibfnamefont {H.}~\bibnamefont {Lin}}, \ and\
  \bibinfo {author} {\bibfnamefont {C.-X.}\ \bibnamefont {Wu}},\ }\bibfield
  {title} {\enquote {\bibinfo {title} {Evolution of prisoner's dilemma
  strategies on scale-free networks},}\ }\href {\doibase
  10.1016/j.physa.2007.06.008} {\bibfield  {journal} {\bibinfo  {journal}
  {Physica A}\ }\textbf {\bibinfo {volume} {385}},\ \bibinfo {pages} {379--384}
  (\bibinfo {year} {2007})}\BibitemShut {NoStop}%
\bibitem [{\citenamefont {Szab\'o}, \citenamefont {Vukov},\ and\ \citenamefont
  {Szolnoki}(2005)}]{PhysRevE.72.047107}%
  \BibitemOpen
  \bibfield  {author} {\bibinfo {author} {\bibfnamefont {G.}~\bibnamefont
  {Szab\'o}}, \bibinfo {author} {\bibfnamefont {J.}~\bibnamefont {Vukov}}, \
  and\ \bibinfo {author} {\bibfnamefont {A.}~\bibnamefont {Szolnoki}},\
  }\bibfield  {title} {\enquote {\bibinfo {title} {Phase diagrams for an
  evolutionary prisoner’s dilemma game on two-dimensional lattices},}\ }\href
  {\doibase 10.1103/PhysRevE.72.047107} {\bibfield  {journal} {\bibinfo
  {journal} {Phys. Rev. E}\ }\textbf {\bibinfo {volume} {72}},\ \bibinfo
  {pages} {047107} (\bibinfo {year} {2005})}\BibitemShut {NoStop}%
\bibitem [{\citenamefont {Rong}, \citenamefont {Yang},\ and\ \citenamefont
  {Wang}(2010)}]{PhysRevE.82.047101}%
  \BibitemOpen
  \bibfield  {author} {\bibinfo {author} {\bibfnamefont {Z.}~\bibnamefont
  {Rong}}, \bibinfo {author} {\bibfnamefont {H.-X.}\ \bibnamefont {Yang}}, \
  and\ \bibinfo {author} {\bibfnamefont {W.-X.}\ \bibnamefont {Wang}},\
  }\bibfield  {title} {\enquote {\bibinfo {title} {Feedback reciprocity
  mechanism promotes the cooperation of highly clustered scale-free
  networks},}\ }\href {\doibase 10.1103/PhysRevE.82.047101} {\bibfield
  {journal} {\bibinfo  {journal} {Phys. Rev. E}\ }\textbf {\bibinfo {volume}
  {82}},\ \bibinfo {pages} {047101} (\bibinfo {year} {2010})}\BibitemShut
  {NoStop}%
\bibitem [{\citenamefont {Rossi}\ and\ \citenamefont {Ahmed}(2015)}]{nr}%
  \BibitemOpen
  \bibfield  {author} {\bibinfo {author} {\bibfnamefont {R.~A.}\ \bibnamefont
  {Rossi}}\ and\ \bibinfo {author} {\bibfnamefont {N.~K.}\ \bibnamefont
  {Ahmed}},\ }\bibfield  {title} {\enquote {\bibinfo {title} {The network data
  repository with interactive graph analytics and visualization},}\ }in\ \href
  {http://networkrepository.com} {\emph {\bibinfo {booktitle} {AAAI'15:
  Proceedings of the Twenty-Ninth AAAI Conference on Artificial
  Intelligence}}}\ (\bibinfo  {publisher} {AAAI Press},\ \bibinfo {year}
  {2015})\ pp.\ \bibinfo {pages} {4292--4293}\BibitemShut {NoStop}%
\bibitem [{\citenamefont {Plischke}\ and\ \citenamefont
  {Bergersen}(2006)}]{plischke2006equilibrium}%
  \BibitemOpen
  \bibfield  {author} {\bibinfo {author} {\bibfnamefont {M.}~\bibnamefont
  {Plischke}}\ and\ \bibinfo {author} {\bibfnamefont {B.}~\bibnamefont
  {Bergersen}},\ }\href@noop {} {\emph {\bibinfo {title} {Equilibrium
  statistical physics}}},\ \bibinfo {edition} {3rd}\ ed.\ (\bibinfo
  {publisher} {World Scientific},\ \bibinfo {year} {2006})\BibitemShut
  {NoStop}%
\bibitem [{\citenamefont {Ising}(1925)}]{Ising}%
  \BibitemOpen
  \bibfield  {author} {\bibinfo {author} {\bibfnamefont {E.}~\bibnamefont
  {Ising}},\ }\bibfield  {title} {\enquote {\bibinfo {title} {Beitrag zur
  {T}heorie des {F}erromagnetismus},}\ }\href {\doibase 10.1007/BF02980577}
  {\bibfield  {journal} {\bibinfo  {journal} {Z. Phys.}\ }\textbf {\bibinfo
  {volume} {31}},\ \bibinfo {pages} {253--258} (\bibinfo {year}
  {1925})}\BibitemShut {NoStop}%
\bibitem [{\citenamefont {Onsager}(1944)}]{Onsager}%
  \BibitemOpen
  \bibfield  {author} {\bibinfo {author} {\bibfnamefont {L.}~\bibnamefont
  {Onsager}},\ }\bibfield  {title} {\enquote {\bibinfo {title} {Crystal
  statistics. {I}. {A} two-dimensional model with an order-disorder
  transition},}\ }\href {\doibase 10.1103/PhysRev.65.117} {\bibfield  {journal}
  {\bibinfo  {journal} {Phys. Rev.}\ }\textbf {\bibinfo {volume} {65}},\
  \bibinfo {pages} {117--149} (\bibinfo {year} {1944})}\BibitemShut {NoStop}%
\bibitem [{\citenamefont {Newman}\ and\ \citenamefont
  {Barkema}(1999)}]{newman1999monte}%
  \BibitemOpen
  \bibfield  {author} {\bibinfo {author} {\bibfnamefont {M.}~\bibnamefont
  {Newman}}\ and\ \bibinfo {author} {\bibfnamefont {G.}~\bibnamefont
  {Barkema}},\ }\href@noop {} {\emph {\bibinfo {title} {{M}onte {C}arlo methods
  in statistical physics}}},\ Vol.~\bibinfo {volume} {24}\ (\bibinfo
  {publisher} {Oxford University Press: New York, USA},\ \bibinfo {year}
  {1999})\BibitemShut {NoStop}%
\bibitem [{\citenamefont {Glauber}(1963)}]{Glauber294}%
  \BibitemOpen
  \bibfield  {author} {\bibinfo {author} {\bibfnamefont {R.~J.}\ \bibnamefont
  {Glauber}},\ }\bibfield  {title} {\enquote {\bibinfo {title}
  {Time‐dependent statistics of the {I}sing model},}\ }\href {\doibase
  10.1063/1.1703954} {\bibfield  {journal} {\bibinfo  {journal} {J. Math.
  Phys.}\ }\textbf {\bibinfo {volume} {4}},\ \bibinfo {pages} {294--307}
  (\bibinfo {year} {1963})}\BibitemShut {NoStop}%
\bibitem [{\citenamefont {Blume}(1993)}]{BLUME1993387}%
  \BibitemOpen
  \bibfield  {author} {\bibinfo {author} {\bibfnamefont {L.~E.}\ \bibnamefont
  {Blume}},\ }\bibfield  {title} {\enquote {\bibinfo {title} {The statistical
  mechanics of strategic interaction},}\ }\href {\doibase
  10.1006/game.1993.1023} {\bibfield  {journal} {\bibinfo  {journal} {Games
  Econ. Behav.}\ }\textbf {\bibinfo {volume} {5}},\ \bibinfo {pages} {387--424}
  (\bibinfo {year} {1993})}\BibitemShut {NoStop}%
\bibitem [{\citenamefont {Fudenberg}\ and\ \citenamefont
  {Levine}(1998)}]{FUDENBERG1998631}%
  \BibitemOpen
  \bibfield  {author} {\bibinfo {author} {\bibfnamefont {D.}~\bibnamefont
  {Fudenberg}}\ and\ \bibinfo {author} {\bibfnamefont {D.}~\bibnamefont
  {Levine}},\ }\bibfield  {title} {\enquote {\bibinfo {title} {Learning in
  games},}\ }\href {\doibase 10.1016/S0014-2921(98)00011-7} {\bibfield
  {journal} {\bibinfo  {journal} {Eur. Econ. Rev.}\ }\textbf {\bibinfo {volume}
  {42}},\ \bibinfo {pages} {631--639} (\bibinfo {year} {1998})}\BibitemShut
  {NoStop}%
\bibitem [{\citenamefont {Monderer}\ and\ \citenamefont
  {Shapley}(1996)}]{MONDERER1996258}%
  \BibitemOpen
  \bibfield  {author} {\bibinfo {author} {\bibfnamefont {D.}~\bibnamefont
  {Monderer}}\ and\ \bibinfo {author} {\bibfnamefont {L.~S.}\ \bibnamefont
  {Shapley}},\ }\bibfield  {title} {\enquote {\bibinfo {title} {Fictitious play
  property for games with identical interests},}\ }\href {\doibase
  10.1006/jeth.1996.0014} {\bibfield  {journal} {\bibinfo  {journal} {J. Econ.
  Theory}\ }\textbf {\bibinfo {volume} {68}},\ \bibinfo {pages} {258--265}
  (\bibinfo {year} {1996})}\BibitemShut {NoStop}%
\bibitem [{\citenamefont {Koopmans}(1951)}]{koopmans1951activity}%
  \BibitemOpen
  \bibfield  {author} {\bibinfo {author} {\bibfnamefont {T.}~\bibnamefont
  {Koopmans}},\ }\href@noop {} {\emph {\bibinfo {title} {Activity analysis of
  production and allocation}}}\ (\bibinfo  {publisher} {Wiley},\ \bibinfo
  {year} {1951})\BibitemShut {NoStop}%
\bibitem [{\citenamefont {Binder}(1981)}]{PhysRevLett.47.693}%
  \BibitemOpen
  \bibfield  {author} {\bibinfo {author} {\bibfnamefont {K.}~\bibnamefont
  {Binder}},\ }\bibfield  {title} {\enquote {\bibinfo {title} {Critical
  properties from {M}onte {C}arlo coarse graining and renormalization},}\
  }\href {\doibase 10.1103/PhysRevLett.47.693} {\bibfield  {journal} {\bibinfo
  {journal} {Phys. Rev. Lett.}\ }\textbf {\bibinfo {volume} {47}},\ \bibinfo
  {pages} {693--696} (\bibinfo {year} {1981})}\BibitemShut {NoStop}%
\bibitem [{\citenamefont {Ferrenberg}\ and\ \citenamefont
  {Landau}(1991)}]{PhysRevB.44.5081}%
  \BibitemOpen
  \bibfield  {author} {\bibinfo {author} {\bibfnamefont {A.~M.}\ \bibnamefont
  {Ferrenberg}}\ and\ \bibinfo {author} {\bibfnamefont {D.~P.}\ \bibnamefont
  {Landau}},\ }\bibfield  {title} {\enquote {\bibinfo {title} {Critical
  behavior of the three-dimensional {I}sing model: A high-resolution {M}onte
  {C}arlo study},}\ }\href {\doibase 10.1103/PhysRevB.44.5081} {\bibfield
  {journal} {\bibinfo  {journal} {Phys. Rev. B}\ }\textbf {\bibinfo {volume}
  {44}},\ \bibinfo {pages} {5081--5091} (\bibinfo {year} {1991})}\BibitemShut
  {NoStop}%
\bibitem [{\citenamefont {Yu}(2017)}]{PhysRevE.95.012101}%
  \BibitemOpen
  \bibfield  {author} {\bibinfo {author} {\bibfnamefont {U.}~\bibnamefont
  {Yu}},\ }\bibfield  {title} {\enquote {\bibinfo {title} {Phase transition in
  the majority-vote model on the {A}rchimedean lattices},}\ }\href {\doibase
  10.1103/PhysRevE.95.012101} {\bibfield  {journal} {\bibinfo  {journal} {Phys.
  Rev. E}\ }\textbf {\bibinfo {volume} {95}},\ \bibinfo {pages} {012101}
  (\bibinfo {year} {2017})}\BibitemShut {NoStop}%
\bibitem [{\citenamefont {Bethe}(1935)}]{Bethe}%
  \BibitemOpen
  \bibfield  {author} {\bibinfo {author} {\bibfnamefont {H.~A.}\ \bibnamefont
  {Bethe}},\ }\bibfield  {title} {\enquote {\bibinfo {title} {Statistical
  theory of superlattices},}\ }\href {\doibase 10.1098/rspa.1935.0122}
  {\bibfield  {journal} {\bibinfo  {journal} {Proc. R. Soc. Lond. A}\ }\textbf
  {\bibinfo {volume} {150}},\ \bibinfo {pages} {552--575} (\bibinfo {year}
  {1935})}\BibitemShut {NoStop}%
\bibitem [{\citenamefont {Peierls}(1936)}]{Peierls}%
  \BibitemOpen
  \bibfield  {author} {\bibinfo {author} {\bibfnamefont {R.}~\bibnamefont
  {Peierls}},\ }\bibfield  {title} {\enquote {\bibinfo {title} {Statistical
  theory of superlattices with unequal concentrations of the components},}\
  }\href {\doibase 10.1098/rspa.1936.0047} {\bibfield  {journal} {\bibinfo
  {journal} {Proc. R. Soc. Lond. A}\ }\textbf {\bibinfo {volume} {154}},\
  \bibinfo {pages} {207--222} (\bibinfo {year} {1936})}\BibitemShut {NoStop}%
\bibitem [{\citenamefont {Barrat}\ and\ \citenamefont {Weigt}(2000)}]{Barrat}%
  \BibitemOpen
  \bibfield  {author} {\bibinfo {author} {\bibfnamefont {A.}~\bibnamefont
  {Barrat}}\ and\ \bibinfo {author} {\bibfnamefont {M.}~\bibnamefont {Weigt}},\
  }\bibfield  {title} {\enquote {\bibinfo {title} {On the properties of
  small-world network models},}\ }\href {\doibase 10.1007/s100510050067}
  {\bibfield  {journal} {\bibinfo  {journal} {Eur. Phys. J. B}\ }\textbf
  {\bibinfo {volume} {13}},\ \bibinfo {pages} {547--560} (\bibinfo {year}
  {2000})}\BibitemShut {NoStop}%
\bibitem [{\citenamefont {Hong}, \citenamefont {Kim},\ and\ \citenamefont
  {Choi}(2002)}]{PhysRevE.66.018101}%
  \BibitemOpen
  \bibfield  {author} {\bibinfo {author} {\bibfnamefont {H.}~\bibnamefont
  {Hong}}, \bibinfo {author} {\bibfnamefont {B.~J.}\ \bibnamefont {Kim}}, \
  and\ \bibinfo {author} {\bibfnamefont {M.~Y.}\ \bibnamefont {Choi}},\
  }\bibfield  {title} {\enquote {\bibinfo {title} {Comment on ``{I}sing model
  on a small world network''},}\ }\href {\doibase 10.1103/PhysRevE.66.018101}
  {\bibfield  {journal} {\bibinfo  {journal} {Phys. Rev. E}\ }\textbf {\bibinfo
  {volume} {66}},\ \bibinfo {pages} {018101} (\bibinfo {year}
  {2002})}\BibitemShut {NoStop}%
\bibitem [{\citenamefont {P\ifmmode~\mbox{\c{e}}\else
  \c{e}\fi{}kalski}(2001)}]{PhysRevE.64.057104}%
  \BibitemOpen
  \bibfield  {author} {\bibinfo {author} {\bibfnamefont {A.}~\bibnamefont
  {P\ifmmode~\mbox{\c{e}}\else \c{e}\fi{}kalski}},\ }\bibfield  {title}
  {\enquote {\bibinfo {title} {{I}sing model on a small world network},}\
  }\href {\doibase 10.1103/PhysRevE.64.057104} {\bibfield  {journal} {\bibinfo
  {journal} {Phys. Rev. E}\ }\textbf {\bibinfo {volume} {64}},\ \bibinfo
  {pages} {057104} (\bibinfo {year} {2001})}\BibitemShut {NoStop}%
\bibitem [{\citenamefont {Grinstein}, \citenamefont {Jayaprakash},\ and\
  \citenamefont {He}(1985)}]{Grinstein}%
  \BibitemOpen
  \bibfield  {author} {\bibinfo {author} {\bibfnamefont {G.}~\bibnamefont
  {Grinstein}}, \bibinfo {author} {\bibfnamefont {C.}~\bibnamefont
  {Jayaprakash}}, \ and\ \bibinfo {author} {\bibfnamefont {Y.}~\bibnamefont
  {He}},\ }\bibfield  {title} {\enquote {\bibinfo {title} {Statistical
  mechanics of probabilistic cellular automata},}\ }\href {\doibase
  10.1103/PhysRevLett.55.2527} {\bibfield  {journal} {\bibinfo  {journal}
  {Phys. Rev. Lett.}\ }\textbf {\bibinfo {volume} {55}},\ \bibinfo {pages}
  {2527--2530} (\bibinfo {year} {1985})}\BibitemShut {NoStop}%
\bibitem [{\citenamefont {Newman}\ and\ \citenamefont
  {Girvan}(2004)}]{PhysRevE.69.026113}%
  \BibitemOpen
  \bibfield  {author} {\bibinfo {author} {\bibfnamefont {M.~E.~J.}\
  \bibnamefont {Newman}}\ and\ \bibinfo {author} {\bibfnamefont
  {M.}~\bibnamefont {Girvan}},\ }\bibfield  {title} {\enquote {\bibinfo {title}
  {Finding and evaluating community structure in networks},}\ }\href {\doibase
  10.1103/PhysRevE.69.026113} {\bibfield  {journal} {\bibinfo  {journal} {Phys.
  Rev. E}\ }\textbf {\bibinfo {volume} {69}},\ \bibinfo {pages} {026113}
  (\bibinfo {year} {2004})}\BibitemShut {NoStop}%
\bibitem [{\citenamefont {Ebel}\ and\ \citenamefont
  {Bornholdt}(2002)}]{PhysRevE.66.056118}%
  \BibitemOpen
  \bibfield  {author} {\bibinfo {author} {\bibfnamefont {H.}~\bibnamefont
  {Ebel}}\ and\ \bibinfo {author} {\bibfnamefont {S.}~\bibnamefont
  {Bornholdt}},\ }\bibfield  {title} {\enquote {\bibinfo {title}
  {Coevolutionary games on networks},}\ }\href {\doibase
  10.1103/PhysRevE.66.056118} {\bibfield  {journal} {\bibinfo  {journal} {Phys.
  Rev. E}\ }\textbf {\bibinfo {volume} {66}},\ \bibinfo {pages} {056118}
  (\bibinfo {year} {2002})}\BibitemShut {NoStop}%
\bibitem [{\citenamefont {Wu}\ and\ \citenamefont
  {Hadzibeganovic}(2020)}]{WU20201}%
  \BibitemOpen
  \bibfield  {author} {\bibinfo {author} {\bibfnamefont {Q.}~\bibnamefont
  {Wu}}\ and\ \bibinfo {author} {\bibfnamefont {T.}~\bibnamefont
  {Hadzibeganovic}},\ }\bibfield  {title} {\enquote {\bibinfo {title} {An
  individual-based modeling framework for infectious disease spreading in
  clustered complex networks},}\ }\href {\doibase 10.1016/j.apm.2020.02.012}
  {\bibfield  {journal} {\bibinfo  {journal} {Appl. Math. Model.}\ }\textbf
  {\bibinfo {volume} {83}},\ \bibinfo {pages} {1--12} (\bibinfo {year}
  {2020})}\BibitemShut {NoStop}%
\bibitem [{\citenamefont {Bianconi}(2002)}]{BIANCONI2002166}%
  \BibitemOpen
  \bibfield  {author} {\bibinfo {author} {\bibfnamefont {G.}~\bibnamefont
  {Bianconi}},\ }\bibfield  {title} {\enquote {\bibinfo {title} {Mean field
  solution of the {I}sing model on a {B}arab\'asi-{A}lbert network},}\ }\href
  {\doibase 10.1016/S0375-9601(02)01232-X} {\bibfield  {journal} {\bibinfo
  {journal} {Phys. Lett. A}\ }\textbf {\bibinfo {volume} {303}},\ \bibinfo
  {pages} {166--168} (\bibinfo {year} {2002})}\BibitemShut {NoStop}%
\bibitem [{\citenamefont {Edwards}\ and\ \citenamefont
  {Anderson}(1975)}]{Edwards_1975}%
  \BibitemOpen
  \bibfield  {author} {\bibinfo {author} {\bibfnamefont {S.~F.}\ \bibnamefont
  {Edwards}}\ and\ \bibinfo {author} {\bibfnamefont {P.~W.}\ \bibnamefont
  {Anderson}},\ }\bibfield  {title} {\enquote {\bibinfo {title} {Theory of spin
  glasses},}\ }\href {\doibase 10.1088/0305-4608/5/5/017} {\bibfield  {journal}
  {\bibinfo  {journal} {J. Phys. F: Met. Phys.}\ }\textbf {\bibinfo {volume}
  {5}},\ \bibinfo {pages} {965--974} (\bibinfo {year} {1975})}\BibitemShut
  {NoStop}%
\bibitem [{\citenamefont {Krawczyk}\ \emph {et~al.}(2005)\citenamefont
  {Krawczyk}, \citenamefont {Malarz}, \citenamefont {Kawecka-Magiera},
  \citenamefont {Maksymowicz},\ and\ \citenamefont
  {Ku\l{}akowski}}]{PhysRevB.72.024445}%
  \BibitemOpen
  \bibfield  {author} {\bibinfo {author} {\bibfnamefont {M.~J.}\ \bibnamefont
  {Krawczyk}}, \bibinfo {author} {\bibfnamefont {K.}~\bibnamefont {Malarz}},
  \bibinfo {author} {\bibfnamefont {B.}~\bibnamefont {Kawecka-Magiera}},
  \bibinfo {author} {\bibfnamefont {A.~Z.}\ \bibnamefont {Maksymowicz}}, \ and\
  \bibinfo {author} {\bibfnamefont {K.}~\bibnamefont {Ku\l{}akowski}},\
  }\bibfield  {title} {\enquote {\bibinfo {title} {Spin-glass properties of an
  {I}sing antiferromagnet on the {A}rchimedean {($3$, $12^{2}$)} lattice},}\
  }\href {\doibase 10.1103/PhysRevB.72.024445} {\bibfield  {journal} {\bibinfo
  {journal} {Phys. Rev. B}\ }\textbf {\bibinfo {volume} {72}},\ \bibinfo
  {pages} {024445} (\bibinfo {year} {2005})}\BibitemShut {NoStop}%
\bibitem [{\citenamefont {Yu}(2015)}]{PhysRevE.91.062121}%
  \BibitemOpen
  \bibfield  {author} {\bibinfo {author} {\bibfnamefont {U.}~\bibnamefont
  {Yu}},\ }\bibfield  {title} {\enquote {\bibinfo {title} {{I}sing
  antiferromagnet on the {A}rchimedean lattices},}\ }\href {\doibase
  10.1103/PhysRevE.91.062121} {\bibfield  {journal} {\bibinfo  {journal} {Phys.
  Rev. E}\ }\textbf {\bibinfo {volume} {91}},\ \bibinfo {pages} {062121}
  (\bibinfo {year} {2015})}\BibitemShut {NoStop}%
\bibitem [{\citenamefont {Voitalov}\ \emph {et~al.}(2019)\citenamefont
  {Voitalov}, \citenamefont {van~der Hoorn}, \citenamefont {van~der Hofstad},\
  and\ \citenamefont {Krioukov}}]{PhysRevResearch.1.033034}%
  \BibitemOpen
  \bibfield  {author} {\bibinfo {author} {\bibfnamefont {I.}~\bibnamefont
  {Voitalov}}, \bibinfo {author} {\bibfnamefont {P.}~\bibnamefont {van~der
  Hoorn}}, \bibinfo {author} {\bibfnamefont {R.}~\bibnamefont {van~der
  Hofstad}}, \ and\ \bibinfo {author} {\bibfnamefont {D.}~\bibnamefont
  {Krioukov}},\ }\bibfield  {title} {\enquote {\bibinfo {title} {Scale-free
  networks well done},}\ }\href {\doibase 10.1103/PhysRevResearch.1.033034}
  {\bibfield  {journal} {\bibinfo  {journal} {Phys. Rev. Research}\ }\textbf
  {\bibinfo {volume} {1}},\ \bibinfo {pages} {033034} (\bibinfo {year}
  {2019})}\BibitemShut {NoStop}%
\bibitem [{\citenamefont {Golosovsky}(2017)}]{PhysRevE.96.032306}%
  \BibitemOpen
  \bibfield  {author} {\bibinfo {author} {\bibfnamefont {M.}~\bibnamefont
  {Golosovsky}},\ }\bibfield  {title} {\enquote {\bibinfo {title} {Power-law
  citation distributions are not scale-free},}\ }\href {\doibase
  10.1103/PhysRevE.96.032306} {\bibfield  {journal} {\bibinfo  {journal} {Phys.
  Rev. E}\ }\textbf {\bibinfo {volume} {96}},\ \bibinfo {pages} {032306}
  (\bibinfo {year} {2017})}\BibitemShut {NoStop}%
\bibitem [{\citenamefont {Tsiotas}(2019)}]{Tsiotas6701}%
  \BibitemOpen
  \bibfield  {author} {\bibinfo {author} {\bibfnamefont {D.}~\bibnamefont
  {Tsiotas}},\ }\bibfield  {title} {\enquote {\bibinfo {title} {Detecting
  different topologies immanent in scale-free networks with the same degree
  distribution},}\ }\href {\doibase 10.1073/pnas.1816842116} {\bibfield
  {journal} {\bibinfo  {journal} {Proc. Natl. Acad. Sci. U.S.A.}\ }\textbf
  {\bibinfo {volume} {116}},\ \bibinfo {pages} {6701--6706} (\bibinfo {year}
  {2019})}\BibitemShut {NoStop}%
\bibitem [{\citenamefont {Clauset}, \citenamefont {Moore},\ and\ \citenamefont
  {Newman}(2008)}]{clauset2008hierarchical}%
  \BibitemOpen
  \bibfield  {author} {\bibinfo {author} {\bibfnamefont {A.}~\bibnamefont
  {Clauset}}, \bibinfo {author} {\bibfnamefont {C.}~\bibnamefont {Moore}}, \
  and\ \bibinfo {author} {\bibfnamefont {M.~E.}\ \bibnamefont {Newman}},\
  }\bibfield  {title} {\enquote {\bibinfo {title} {Hierarchical structure and
  the prediction of missing links in networks},}\ }\href@noop {} {\bibfield
  {journal} {\bibinfo  {journal} {Nature}\ }\textbf {\bibinfo {volume} {453}},\
  \bibinfo {pages} {98--101} (\bibinfo {year} {2008})}\BibitemShut {NoStop}%
\end{thebibliography}%

\end{document}